\documentclass[preprint]{aastex}
\usepackage{subfigure}

\slugcomment{Draft version - \today}
\begin{document}

\title{Strong [\ion{O}{3}] Objects Among SDSS Broad-Line Active Galaxies}
\author{Randi R. Ludwig\altaffilmark{1}}
\author{Beverley Wills\altaffilmark{1}}
\author{Jenny E. Greene\altaffilmark{1, 2}}
\author{Edward L. Robinson\altaffilmark{1}}
\affil{$^1$University of Texas at Austin, Department of Astronomy, 1 University Station, C1400
Austin, TX   78712, USA}
\affil{$^2$Department of Astrophysical Sciences, Princeton University,
Princeton, NJ 08544; Hubble, Princeton-Carnegie Fellow}
\shorttitle{Strong [\ion{O}{3}] in Broad-line AGN}
\shortauthors{LUDWIG et al.}

\email{randi@astro.as.utexas.edu}

\begin{abstract}

We present the results of a spectral principal component analysis on 9046 broad-line AGN from the Sloan Digital Sky Survey.  We examine correlations between spectral regions within various eigenspectra (e.g., between Fe II strength and H$\beta$ width) and confirm that the same trends are 
apparent in spectral measurements, as validation of our technique.  Because we found that our sample had a large range in the equivalent width of [\ion{O}{3}] $\lambda5007$, we divided the data into three subsets based on [\ion{O}{3}] strength.  Of these, only in the sample with the weakest equivalent width of [\ion{O}{3}] were we able to recover the known correlation between [\ion{O}{3}] strength and full width at half maximum of H$\beta$ and their anticorrelation with \ion{Fe}{2} strength.  At the low luminosities considered here ($L_{5100 \mbox{\AA}}$ of $10^{42}-10^{46}$ erg s$^{-1}$), interpretation of the principal components is considerably complicated particularly because of the wide range in [\ion{O}{3}] equivalent width.  We speculate that variations in covering factor are responsible for this wide range in [\ion{O}{3}] strength.


\end{abstract}

\section{Introduction}
\label{sec:intro}

Active galactic nuclei (AGN) display an incredibly diverse set of observed properties.  They reside in host galaxies of many morphologies, and have a range of radio luminosities and x-ray spectra. There are obscured AGN, known as Type II AGN, that exhibit only narrow forbidden emission lines emerging from the narrow line region (NLR), with scales $\gtrsim$ 100 pcs.  Also, there are broad-line AGN that exhibit both NLR emission and broad ($>1000$ km s$^{-1}$) emission lines from the broad line region (BLR) with scales of light days.  There is great variation in the strengths, shapes, and widths of emission lines. This diversity provides many opportunities for investigating the physics at work in these systems.

A powerful tool for studying AGN phenomenology is principal component analysis (PCA).  PCA finds the eigenvectors, or most dominant relationships, within a set of input data.  \citet{1992ApJS...80..109B}, hereafter BG92, performed a principal component analysis on a sample of $z < 0.5$ quasars from the Palomar-Green Bright Quasar Survey \citep[BQS; ][]{1983ApJ...269..352S}, a UV-selected sample of point-like quasars (where most of the light fell within 1" on blue Palomar Sky Survey prints).   Spectral parameters that they used in the PCA included the the EW and FWHM of H$\beta$, optical \ion{Fe}{2}, [\ion{O}{3}] $\lambda$5007, and He II $\lambda$4686;  ratios of the EWs of optical \ion{Fe}{2}, [\ion{O}{3}], and He II relative to H$\beta$; and magnitude and peak flux of [\ion{O}{3}] relative to H$\beta$, absolute magnitude, optical-to-X-ray spectral slope ($\alpha_{\rm{ox}}$), radio-loudness, and line shift, shape, and asymmetry of H$\beta$.  The eigenvector that distinguished most quasars from one another was dominated by the anticorrelation between the broad-line \ion{Fe}{2} strength relative to broad H$\beta$ versus the peak flux of [\ion{O}{3}].  They found a second eigenvector that included a correlation among optical luminosity $L$, He II, and $\alpha_{\rm{ox}}$.  

The BG92 anticorrelation between \ion{Fe}{2}/H$\beta$ and [\ion{O}{3}]/H$\beta$, called eigenvector 1 (EV1), revealed a tie between the BLR and NLR, even though these regions occupy very different physical scales.  At one end of this relationship lie narrow-line Seyfert 1 galaxies \citep[NLS1; ][]{1985ApJ...297..166O}, with relatively narrow broad lines, strong \ion{Fe}{2}/H$\beta$, weak [\ion{O}{3}]/H$\beta$, quiet radio emission, and a steep slope between the x-ray and optical continuum \citep{1994ApJ...420..110L, 1997ApJ...477...93L}.  At the other end of EV1 lie radio galaxies with broad ($> 4000$ km s$^{-1}$) BLR emission, strong [\ion{O}{3}]/H$\beta$, very little \ion{Fe}{2}, and flat slopes between the optical and x-ray regions.  

EV1 includes a correlation with the linewidth of H$\beta$ \citep{1996A&A...305...53B}, which is used to estimate virial black hole masses.  This led to the idea that EV1 could be driven by black hole mass ($M_{\rm{BH}}$) or Eddington ratio \citep[$L_{\rm{bol}}/L_{\rm{Edd}}$; BG92, ][]{1994ApJ...420..110L, 1997ApJ...477...93L}, so that high accretion ratios correspond to the NLS1 end of EV1. A correlation with $L_{\rm{bol}}/L_{\rm{Edd}}$ provides an explanation for the observed x-ray properties in AGN such that a higher accretion state, or higher $L_{\rm{bol}}/L_{\rm{Edd}}$, causes the accretion disk to become thicker and produce more soft x-rays \citep[e.g.][]{2004AJ....127.1799G}.  Physically, this thicker disk could be related to disk outflows that additionally excite \ion{Fe}{2} \citep{2000NewAR..44..531C}.   It is also possible that the covering factor is higher for the BLR in strong \ion{Fe}{2} objects, resulting in less continuum reaching the NLR gas, which would explain the \ion{Fe}{2} - [\ion{O}{3}] anticorrelation.  

Over time, many parameters have been examined for correlations with EV1, such as the slope of the x-ray spectrum \citep{1994ApJ...420..110L,1997ApJ...477...93L,1995MNRAS.277L...5P}, black hole mass and $L_{\rm{bol}}/L_{\rm{Edd}}$ or $L/M$ \citep{1994ApJ...420..110L,1997ApJ...477...93L,1992ApJS...80..109B,2002ApJ...565...78B,2003MNRAS.345.1133M}, radio properties \citep{2002ApJ...565...78B} and the \ion{Fe}{2} emission \citep{2003ApJ...586...52S}.  Also, many authors have utilized PCA to analyze correlations among measured quantities \citep{2004AJ....127.1799G, 2008ApJ...678...22H}, or used PCA to find and investigate outliers with unusual properties \citep{1999ASPC..162..363F}.  Still others have chosen to apply a PCA directly to the spectra \citep{1994ApJ...430..495B, 2003ApJ...586...52S, 2004AJ....128.2603Y}.  These investigations have provided important insights, but we still lack a complete physical understanding of the connection between the inner BLR and the far-ranging NLR.  

The Sloan Digital Sky Survey \citep[SDSS; ][]{2000AJ....120.1579Y} has provided an immense data set of broad-line AGN spectra that can be useful for understanding whether the BG92 EV1 relationships persist in large samples.  The SDSS is especially appropriate:  large samples with well-defined selection criteria, larger luminosity ranges with reasonable completeness, and better spectral resolution than has been available for most broad-line AGN samples in the past.  The sheer number of objects also provides a greater diversity of broad-line AGN properties available for study.  From such a large sample, we can also identify extreme objects for further study.  In this paper we report an SPCA of broad-line AGN from the SDSS.

In \S ~\ref{sec:method} we discuss our sample and SPCA methodology.  In \S ~\ref{sec:results} we present the results of performing spectral PCA on our data set, and find that, over the broad range of properties in the sample, we cannot find simple physical interpretations for the principal components, perhaps because of nonlinear relationships between the physical parameters and the observables.  We further probe the relationships in the sample by dividing it into three subsets.  In \S ~\ref{sec:bg92} we investigate the differences between our work and that of BG92.  In \S ~\ref{sec:outliers}, we discuss properties of our most extreme subset, the strong NLR objects, and examine possible reasons for their extraordinarily strong [\ion{O}{3}] emission.  We find that high covering factors could be a plausible explanation for the strong NLR emission.  We present a summary of our results in \S ~\ref{sec:summary}.

All luminosities in this paper were calculated using the cosmological parameters $H_{0}=70$ km s$^{-1}$ Mpc$^{-1}$, $\Omega_{\rm{M}}=0.3$, and $\Omega_{\Lambda}=0.7$.

\section{Methodology}
\label{sec:method}
\subsection{Sample Definition and Measurements}

We selected a large sample of broad-line AGN spectra from the SDSS Data Release 5 \citep[DR5; ][]{2007ApJS..172..634A,2000AJ....120.1579Y,2002AJ....123..485S}.   The AGN in SDSS are selected by two primary methods.  First, quasar candidates were targeted for spectroscopy by a photometric color and morphological selection above a limiting magnitude of $i < 19.1$ \citep{2002AJ....123.2945R}.  Second, AGN were spectroscopically discovered in spectra that were targeted for other reasons, for instance as probable galaxies \citep{2002AJ....124.1810S, 2001AJ....122.2267E}.  When the latter spectra were cross-correlated with the SDSS quasar template spectrum, they were identified as quasars due to their prominent broad lines and blue continuua.
		
We obtained our sample and spectral measurements from \cite{2007ApJ...662..131S}, which they have since extended to DR5.  We selected all the DR5 objects with spectral coverage of the rest-frame region 4000--6000 \AA{} on which we performed the SPCA.   This selection includes \ion{Fe}{2} and [\ion{O}{3}] in the observed wavelength range of SDSS.  We chose not to extend this wavelength region further to the red because it would result in a more severe redshift restriction, nor to the blue because we wanted the ability to do a more straightforward comparison with BG92 and they did not use data blueward of 4000 \AA{}{} for their sample.  This limited our sample to the redshift range $z \lesssim 0.56$.  We also chose to exclude objects with $z < 0.1$ to reduce the need of fitting the host galaxy continuua in low luminosity AGN.  There are 12040 quasars in SDSS DR5 with $0.1 < z \lesssim 0.56$\footnote{We made no cuts based on the SDSS confidence in the measured redshifts.  After inspecting a random subsample of objects with confidences $< 90\%$, we confirmed that at least 90\% of these are noisy AGN and are not concerned that they will affect our results.}.  For consistency, all redshifts used in this paper are the SDSS DR5 spectroscopic redshifts derived from template matching rather than the Salviander et al. redshifts derived from line fitting. 
		
Salviander et al. fitted the spectra in the wavelength region 4000--5500 \AA{}{} with a power law, \ion{Fe}{2} pseudocontinuum derived from the \cite{2003ApJS..145..199M} template, the narrow-line [\ion{O}{3}] $\lambda\lambda$4960, 5008, and narrow and broad H$\beta$.  Note that there was no fit to the host galaxy continuum.  The [\ion{O}{3}] line ratios were fixed to their laboratory value of 1:3 for the [\ion{O}{3}] $\lambda$4960 and $\lambda$5008 lines respectively, and all wavelengths were fixed to their vacuum values as well.  The narrow component of H$\beta$ was fixed to the same redshift and line profile as the [\ion{O}{3}] line, with 10\% of the flux.  The broad component of H$\beta$ was allowed to vary independently.  The emission lines were all fitted using Gauss-Hermite polynomials, which allow straightforward parameterization of line shapes and asymmetries.   The measurements include $L_{5100 \mbox{\AA}}$, which is $\lambda L_{\lambda}$ at 5100 \AA{}, rest-frame EW$_{\rm{Fe II}}$, and the line-shape, rest-frame EW, integrated line flux, and FWHM for each of the emission lines.  The fitting procedure is described in more detail in Salviander et al.
	
We decided to be more conservative in our sample selection than Salviander et al. by excluding potential Type II AGN from our sample, since we wished to investigate relationships between the BLR and NLR.  To do this, we conservatively eliminated objects with FWHM of broad H$\beta < 2000$ km s$^{-1}$.  We compared the results of performing a spectral principal component analysis (SPCA) on this sample with a sample including objects with FWHM H$\beta$ as low as 1050 km s$^{-1}$ and found that the principal components were virtually identical.  The linewidth cut reduced the sample to 9362 objects.

Lastly, like Salviander et al., we eliminated the objects with a failed fit of the H$\beta$ broad line since we wished to use H$\beta$ parameters in our analyses.  In general, the fits failed when the spectrum had low S/N, absence of the broad line, or cosmetic defects such as cosmic rays.  However, unlike Salviander et al. we chose to include objects with failed fits of [\ion{O}{3}] since they were likely weak [\ion{O}{3}] objects.  The H$\beta$ failed fits removed only 316 objects, or 3\% of the sample.  Our final data set consisted of 9046 objects.  

The objects in our sample were targeted for spectroscopic followup by the SDSS pipeline for a variety of reasons.  We found that 88\% of the sample were primarily targeted as quasars of some sort.  The very low luminosity objects were generally targeted as galaxies, but this represents only 0.4\% of the total sample.  Only 4\% of our sample were ``serendipitous,'' but these do not have unusual principal component properties (see below).  The other 7.6\% were targeted primarily for an assortment of other reasons, including ROSAT and FIRST detections, blue stars, etc.

\subsection{SPCA Procedure}

PCA is a powerful tool because it provides a means of classifying and potentially distilling important relationships among the data \citep[][and references therein]{1992ApJ...398..476F}.  PCA extracts orthogonal eigenvectors, each consisting of a linear combination of input variables.  They are conventionally ordered according to the fraction of sample variance represented by each eigenvector.  Although there is no a priori reason, the hope is that the sample variance can be represented by only a small number of eigenvectors, so PCA results in a drastic simplification of the entire dataset.  For example, BG92 found that 51\% of the variance in their sample was represented by the first two eigenvectors.  Additionally, one hopes that these eigenvectors illuminate physical relationships.  Indeed, BG92 \citep[updated by][]{2002ApJ...565...78B} found that EV1 is related to $M_{BH}$ or Eddington ratio and EV2 is related to $L$.  However, in nature the relationships among various parameters of an object may not be well-described by a linear analysis.  For instance, a certain characteristic may apply to only a subset of objects.  This may make physical interpretations difficult and confuse relationships among the input parameters, but the PCA still appears to provide a useful description of the data.  

While many authors have used direct line and continuum measurements as input data for PCA, there is another approach.  Spectral PCA or SPCA has been proven to reduce various samples \citep{1994ApJ...430..495B, 2003ApJ...586...52S, 2004AJ....128.2603Y} down to only a handful of components that reproduce the vast majority of quasar-to-quasar spectral differences. SPCA uses the flux densities binned in wavelength as input parameters to solve for orthogonal eigenvectors, or principal components \citep[PCs][]{1992ApJ...398..476F}.  Since the input variables are not the same, the features responsible for most of the variance within a sample can be different than in a PCA.  For example, many continuum bins are likely to result  in a greater contribution from the continuum to the PCs in an SPCA.  The resulting PCs are not typical spectra, with flux versus wavelength, but are representations of how wavelength regions correlate with one another ($C_{\lambda}$).  Positive correlations between wavelength regions appear as features pointing the same direction, while anticorrelations appear with opposite directions.  SPCA retains information for each input spectrum in the form of the coefficients (scores) for each PC.  The coefficients can then be compared with various measured properties of the AGN (e.g. black hole masses) in the hopes of gaining additional insights.  One benefit of SPCA is the ability to easily identify extreme objects as outliers in a PC-PC plane using the PC coefficients for each object.  Another benefit of SPCA is that there is no need to extract measured quantities from the spectra, so there are no added uncertainties from continuum estimation or fitting errors.  In our case, the spectrum-to-spectrum noise in the results are reduced by the large numbers of SDSS spectra.  We can include spectra with low signal-to-noise ratios whose features may be difficult to measure accurately, and these spectra will not bias the results.  

While SPCA is a different approach to a PCA like BG92's because it uses very different input parameters, in the case of optical spectra of luminous quasars they still yield meaningful and comparable results.  \citet{2003ApJ...586...52S} carried out an SPCA on the same limited optical region that we used for a sample made up of a subset of the quasars in BG92.  By comparing their components with measured spectral parameters, they were able to justify their physically meaningful interpretations of the PCs.  They also showed that, in the same optical region as in BG92, as few as two PCs dominate the sample variance.  Moreover these, PC1 and PC3, appeared to represent the same physical quantities as BG92's EV1 and EV2.  They showed this by direct correlation of the coefficients of their PCs for each spectrum with the physical quantities apparently underlying the BG92 eigenvectors.  These earlier studies were for relatively small samples of luminous quasars (M$_{V} \lesssim -23$).  Can we extend these studies of the optical region to the much larger SDSS sample, covering a wider range of $L$?	
	
\section{SPCA Results}
\label{sec:results}

We performed an SPCA on our entire sample of 9046 SDSS broad-line AGN, using the code developed by \cite{1992ApJ...398..476F}.  We normalized each spectrum by its mean flux, and we do not divide by the standard deviation.	 In what follows we look for features within the PCs and then confirm that they correspond to trends in the measured properties of the input spectra.

\subsection{Entire Data Set}
	
\begin{figure}[!tp]
\centering
\includegraphics[scale=0.5]{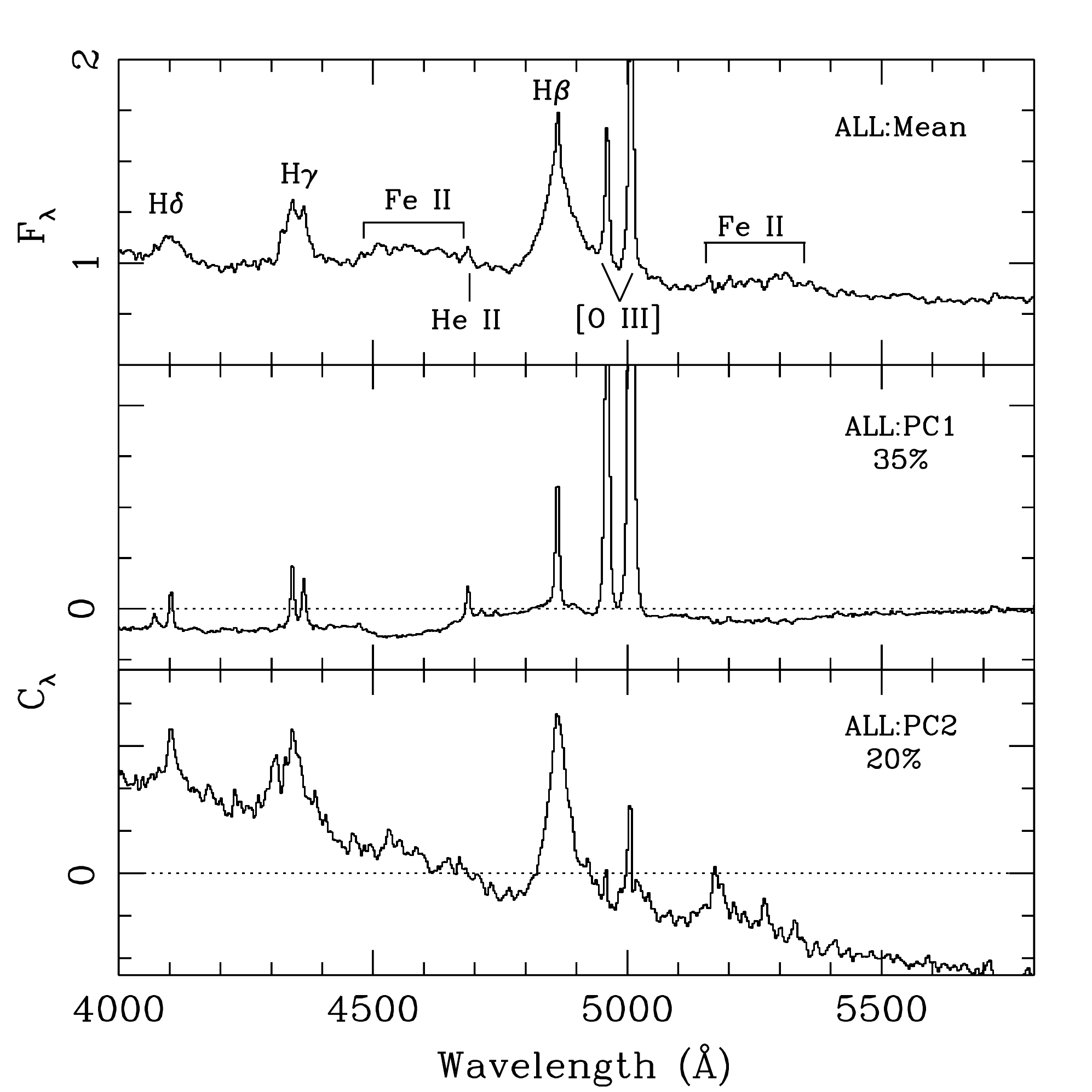}
\caption{Mean spectrum ({\it top}), PC1 ({\it middle}), and PC2 ({\it bottom}) for the entire data set.  ALL:PC1 shows correlations among the NLR lines and ALL:PC2 incorporates the continuum slope and the broad Balmer lines. The dotted line denotes $C_{\lambda} =0$, which distinguishes positive and negative features.}
\label{datasetcomp}
\end{figure}

\begin{figure*}[!tp]
\centering
\subfigure{
\includegraphics[scale=0.35]{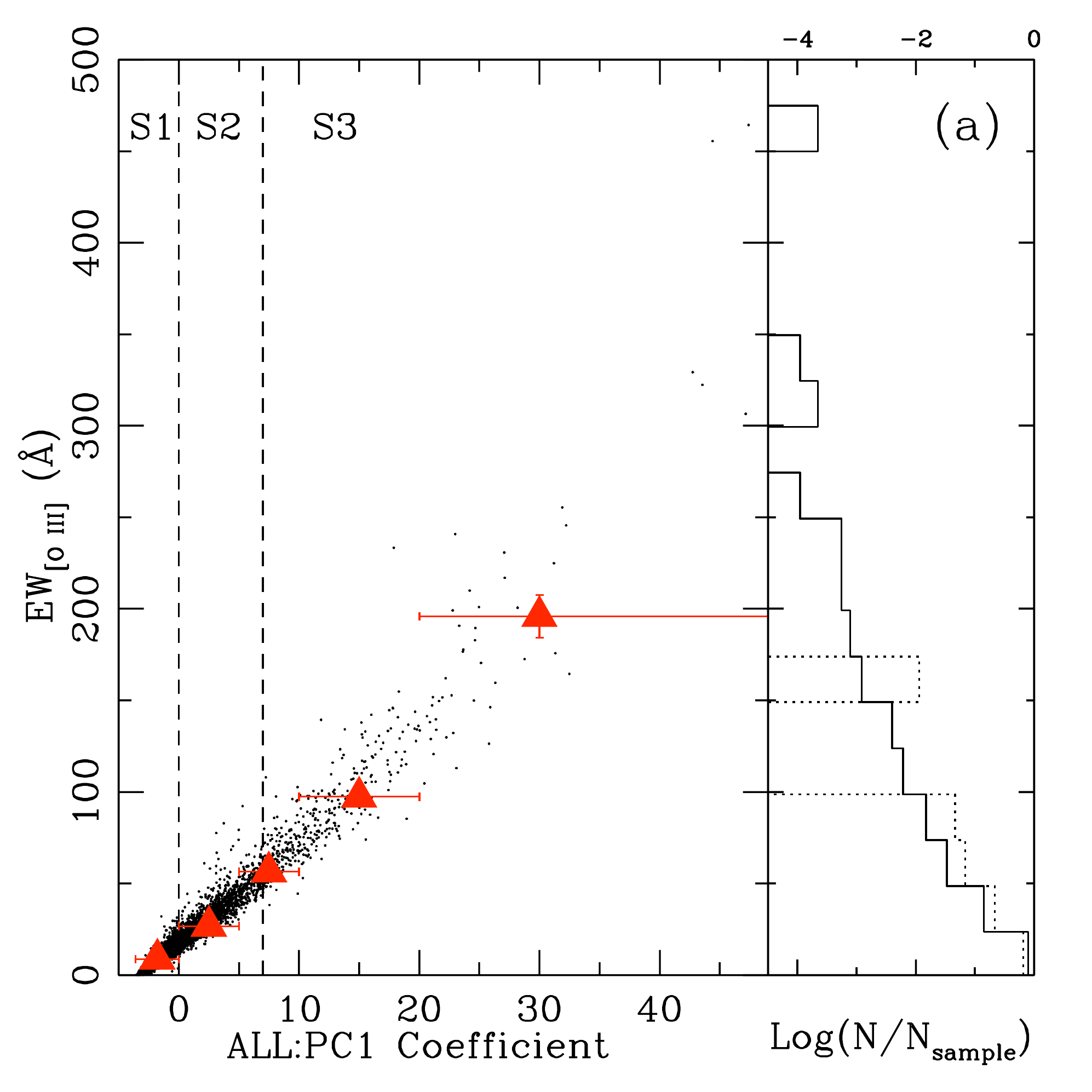}
\label{o3allpc1}
}
\subfigure{
\includegraphics[scale=0.35]{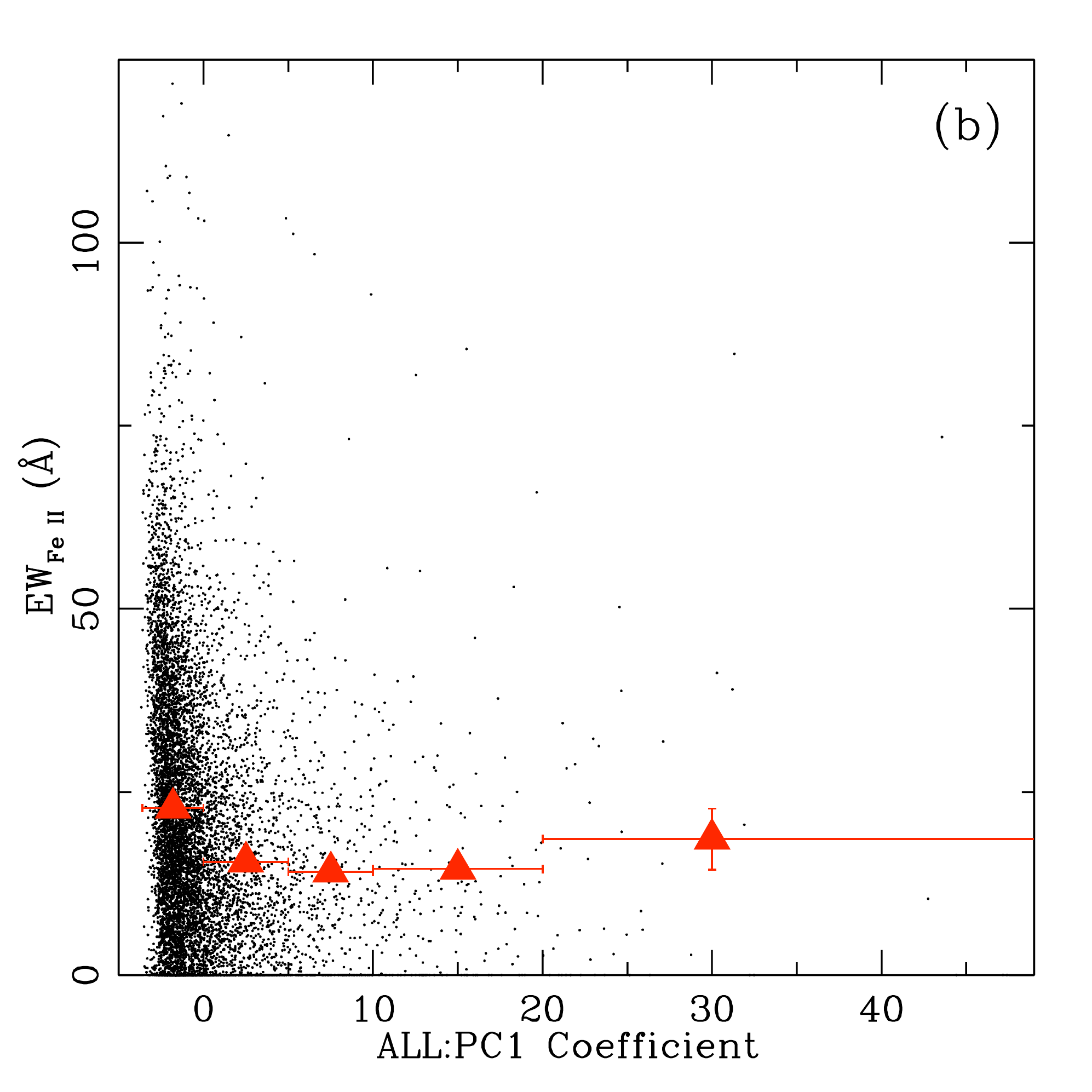}
\label{fe2allpc1}
}
\subfigure{
\includegraphics[scale=0.35]{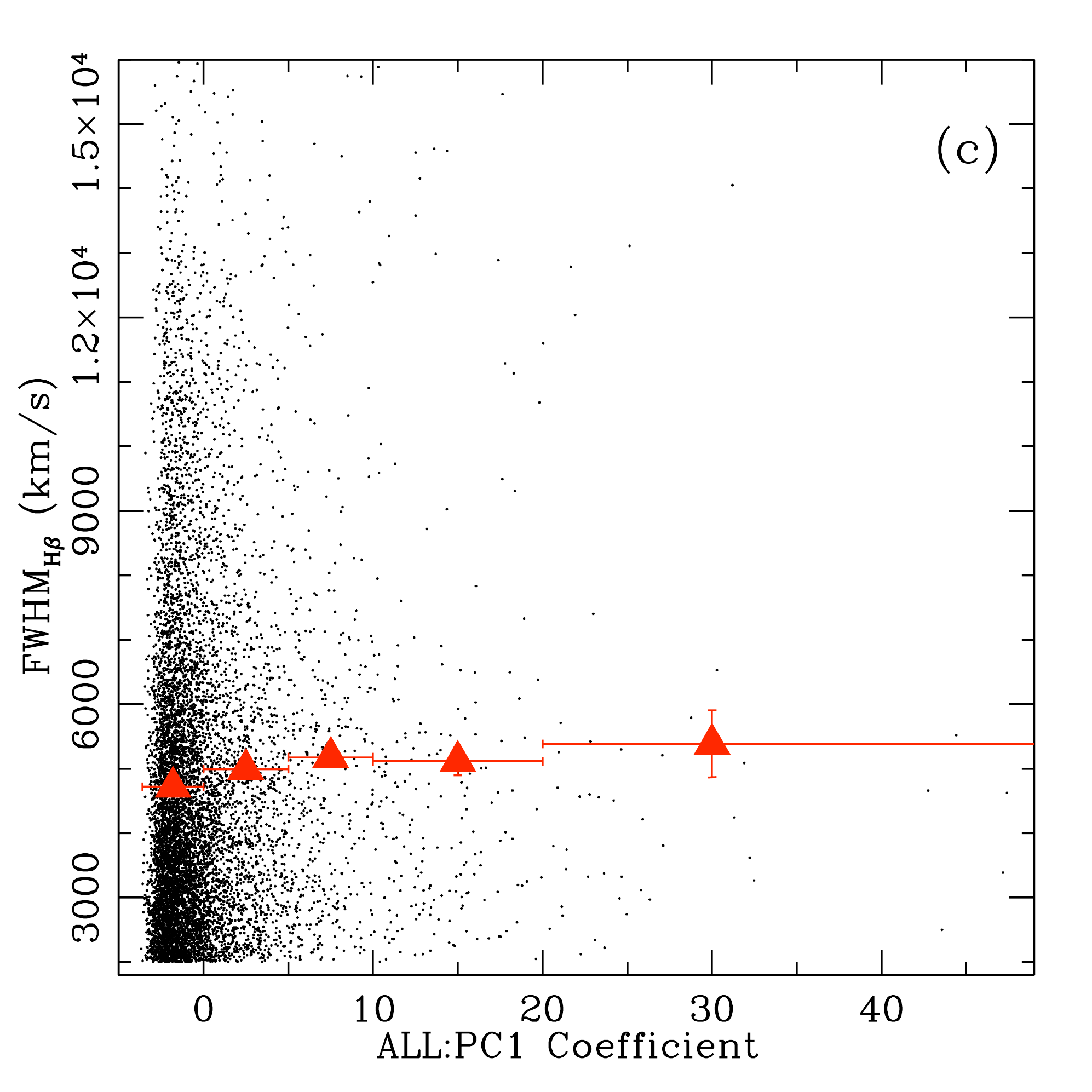}
\label{hbwallpc1}
}
\subfigure{
\includegraphics[scale=0.35]{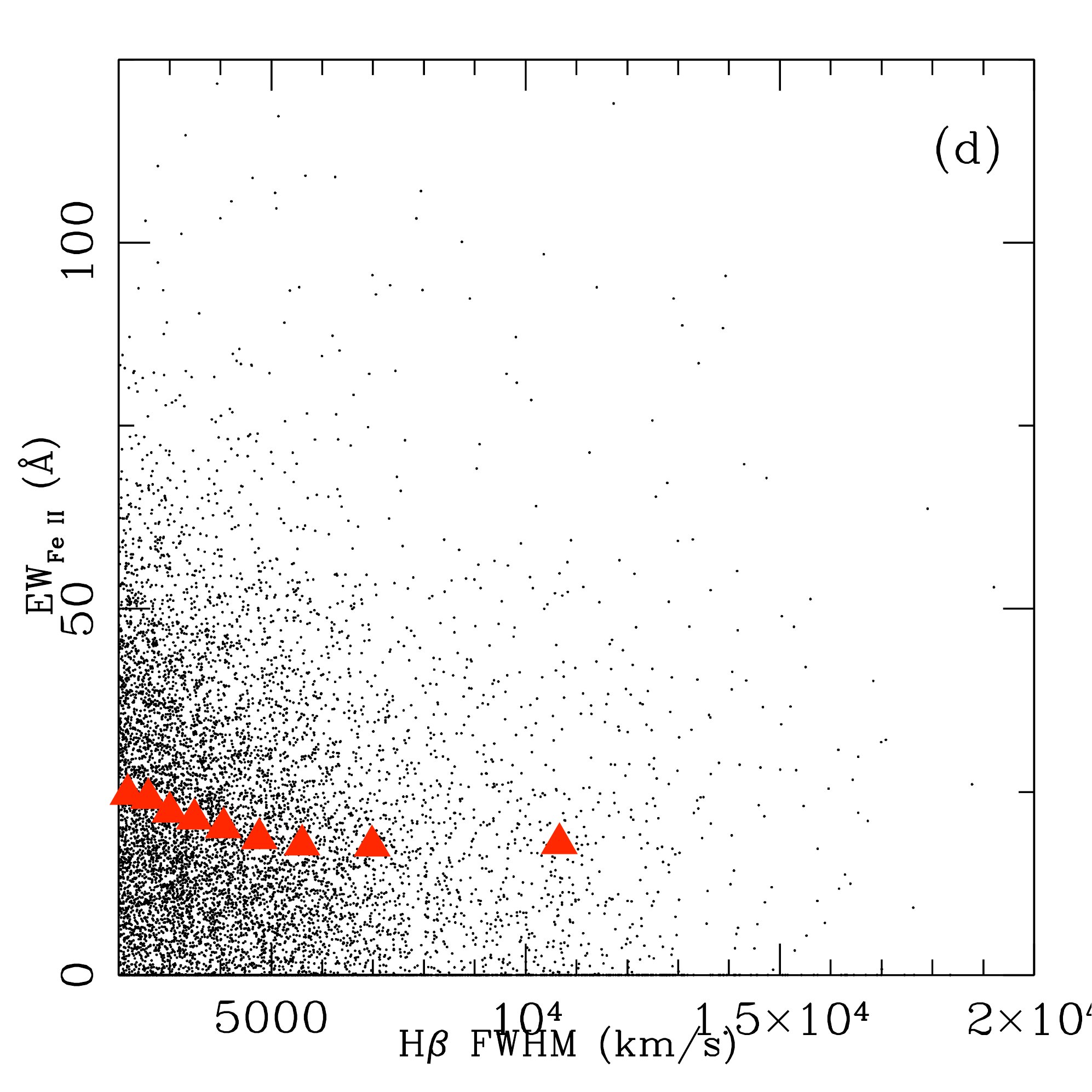}
\label{finalfe2hb}
}
\caption{\textit{(a)} EW$_{\mathrm{[O III]}}$ versus ALL:PC1 Coefficients.  The red triangles represent the mean of the data binned in the x-axis, so that each bin included at least 70 points, where error bars are the bin widths and standard deviations divided by $\sqrt{n-1}$ for each bin.  The error bars are sometimes smaller than the symbols.  We additionally denote subset divisions with the vertical dashed lines, and in the side histogram we show the BG92 distribution in EW$_{\mathrm{[O III]}}$ as the dotted histogram, compared with our objects in the solid histogram. 
\textit{(b)} EW$_{\mathrm{Fe II}}$ versus ALL:PC1 Coefficients.  Symbols are as in \textit{(a)}.
\textit{(c)} FWHM$_{\mathrm{H}\beta}$ versus ALL:PC1 Coefficients.  Symbols are as in \textit{(a)}.
\textit{(d)} shows EW$_{\mathrm{Fe II}}$ versus FWHM$_{\mathrm{H}\beta}$ for the entire sample binned in FWHM$_{\mathrm{H}\beta}$ where error bars are standard deviations divided by $\sqrt{n-1}$. 
\label{fig:paramallpc1}}
\end{figure*}

The mean spectrum and first two PCs for our entire data set are shown in Figure ~\ref{datasetcomp}.  We use ALL:PC1 to denote the first PC from the SPCA run on the entire data set, ALL:PC2 for the second, and so on.  ALL:PC1, which accounts for 35\% of the variance, appears to show a strong correlation among EWs of lines from the NLR, including [\ion{O}{3}], the narrow Balmer lines, and He II $\lambda$4686.  There is a less prominent anticorrelation between the EW of the narrow lines and the EW of the \ion{Fe}{2} blends that originate in the BLR.  If this qualitative interpretation of the PCs is correct, one would expect to see a correlation between the ALL:PC1 coefficients for individual spectra when compared with the measured values of EW$_{\mathrm{[O III]}}$ and an inverse correlation with EW$_{\mathrm{Fe II}}$.  We plot these values in Figure~\ref{fig:paramallpc1} and find that, in fact, the EW$_{\mathrm{[O III]}}$ is indeed strongly correlated with ALL:PC1 coefficients, with a Spearman correlation coefficient of $0.95$.  Figure ~\ref{fe2allpc1} shows a less obvious anticorrelation between EW$_{\mathrm{Fe II}}$ and ALL:PC1, which has a Spearman correlation coefficient of $-0.297$.  Given these relationships, ALL:PC1 is the component that most closely reproduces the correlations in the traditional EV1.  However, in at least one respect ALL:PC1 deviates significantly from the BG92 EV1.  ALL:PC1 does not appear to include any correlation with the linewidth of broad H$\beta$, which would be present in the PC as a ``M''-shaped pattern centered at H$\beta$ \citep[][hereafter S03]{2003ApJ...586...52S}.  There is no strong correlation between ALL:PC1 and the FWHM of H$\beta$ (Spearman correlation coefficient of $0.076$; Figure ~\ref{hbwallpc1}).  At face value, our result disagrees with the classic BG92 finding that the strength of the \ion{Fe}{2} emission strongly anticorrelates with the FWHM of H$\beta$.  By plotting the EW$_{\mathrm{Fe II}}$ directly against the FWHM of H$\beta$ (Figure ~\ref{finalfe2hb}), we note that the entire sample appears to have an anticorrelation between the two, but there is a large amount of scatter (Spearman correlation coefficient of $-0.198$).  Several authors \citep[S03,][]{2004AJ....128.2603Y} have performed similar SPCAs using the rest-frame optical region of the spectrum.  While the small, high-luminosity sample used by S03 seems to recover EV1 correlations, the large samples used here and in \cite{2004AJ....128.2603Y} do not include the strong linewidth dependence of broad H$\beta$.  

The dominant correlations present in ALL:PC2, which accounts for 20\% of the variance, are positive correlations between all broad lines and  continuum shape (objects with a strongly sloped continuum will have more contribution from PC2 compared with objects with a flat continuum).  We also find that ALL:PC2 is correlated with $L_{\rm{5100}}$, with a Spearman correlation probability of $< 0.001$ that they are not correlated.  Further components of the entire data set each account for $< 12$\% of the variance among the spectra and we do not offer interpretations of them.

Since our large sample and others do not seem to entirely recover the traditional EV1 correlations, we wonder if our choice of such a large, diverse sample is potentially masking our ability to notice relationships among the data.  Over the range of properties that the objects in our sample exhibit, the distribution of a given property may saturate for a subset of objects.  Alternatively, the relationship between properties need not be linear.  For instance, the EW$_{\mathrm{[O III]}}$ spans such a large range relative to other properties that it dominates ALL:PC1 and therefore the SPCA results of the entire data set.  Thus, an SPCA that uses the entire sample is possibly less likely to result in PCs that are directly interpretable in terms of physical properties.

\subsection{Subsets}

\begin{figure}[!tp]
\centering
\includegraphics[scale=0.5]{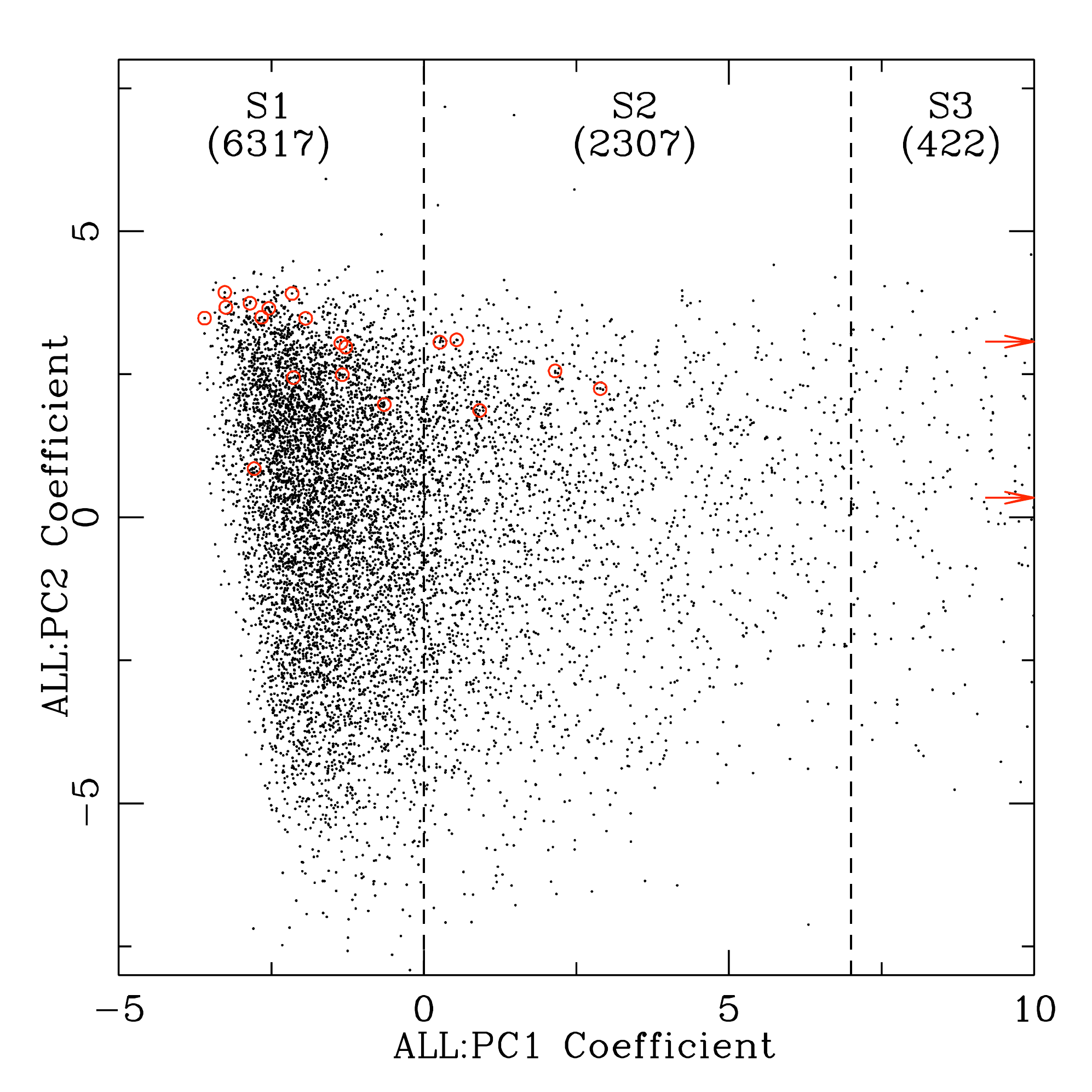}
\caption{ALL:PC2 coefficients vs. ALL:PC1 coefficients.  The dashed lines at ALL:PC1 $=0$ and ALL:PC1 $=  7$ denote our subset divisions with the number of objects in each subset in parentheses.  There are ~200 objects at higher values of ALL:PC1, not shown, with values ranging up to ALL:PC1 $= 60$.  The red circles denote the 21 PG quasars with SDSS spectroscopy, discussed in section \S ~\ref{sec:bg92}.  There are two PG quasars in S3 at ALL:PC1 values of 15 and 33, denoted by the red arrows.}
\label{pcvpc}
\end{figure}

One way to investigate relationships that might be diluted by the large range of properties in the entire sample is to divide the sample into subsets.  SDSS, with the large sample it provides, affords us an opportunity to investigate subsets with particular properties.  Because ALL:PC1 encompasses the relationships representing the largest spectrum-to-spectrum variation among the objects in the entire data set, we chose to divide our sample into subsets based on the ALL:PC1 coefficient of each object.  We settled on the following three subsets, which isolate the extremes of the EW$_{\mathrm{[O III]}}$ range:  S1, which has ALL:PC1~$< 0$~and includes 6317 objects, S2 with $0 <$~ALL:PC1~$< 7$ includes 2307 objects, and S3 with~ALL:PC1 $> 7$ includes only 422 objects (see Figure ~\ref{pcvpc}).  These subset divisions are roughly equivalent to EW$_{\mathrm{[O III]}} < 15$~\AA{}, $15-50$~\AA{}, and $> 50$~\AA{} for S1, S2, and S3, respectively.  However, by restricting the range of EW$_{\mathrm{[O III]}}$, we realize we might be reducing the importance of any relationships that do include EW$_{\mathrm{[O III]}}$.

\begin{figure*}
\centering
\includegraphics[scale=0.9]{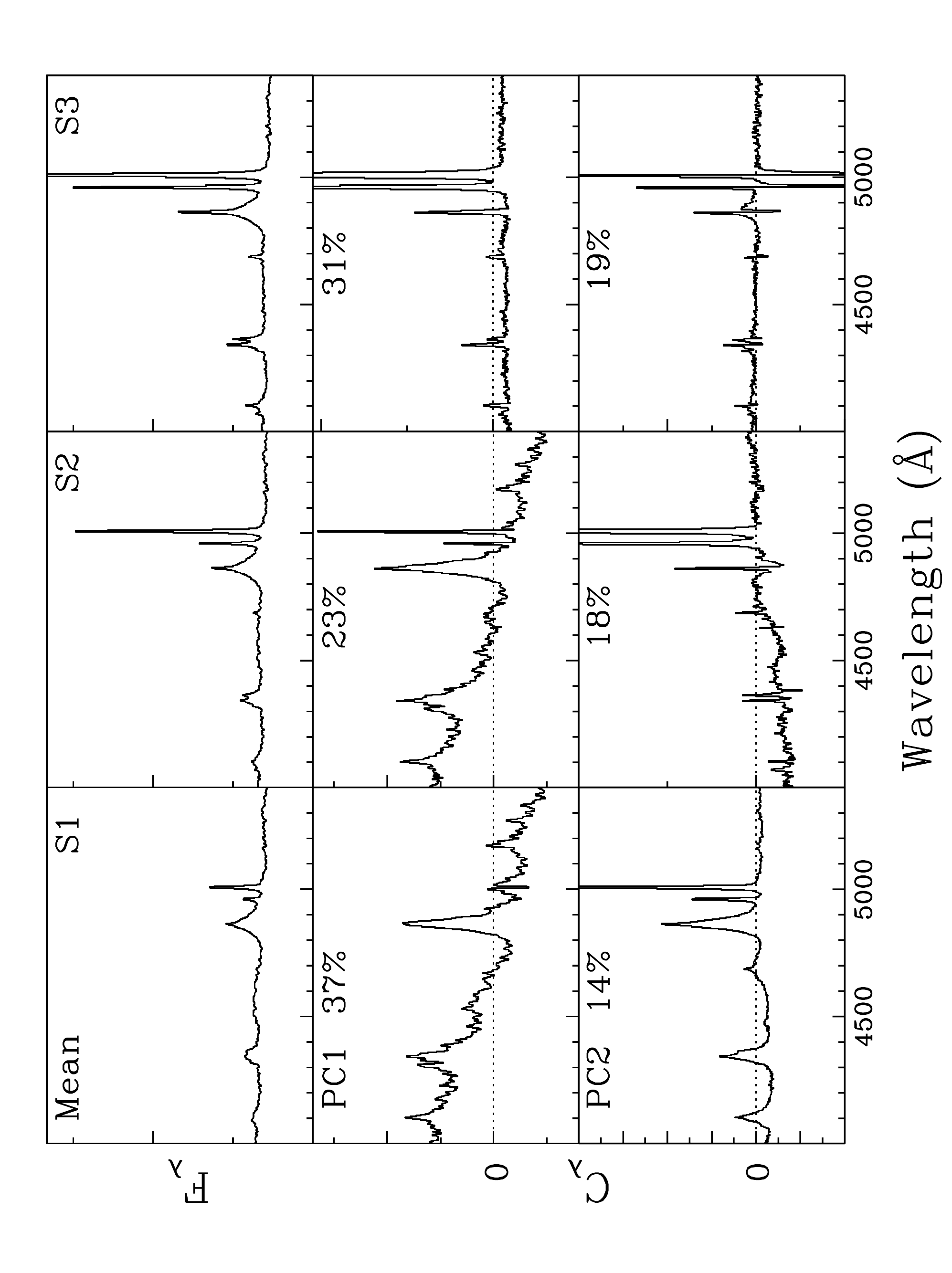}
caption{Mean spectrum, new PC1 and new PC2 for all three subsets, where the left column contains results for S1 (ALL:PC1 $< 0$), the middle column contains S2 results ($0 < $ALL:PC1$ < 7$), and the right column contains S3 results (ALL:PC1$> 7$).  In the mean spectrum, notice the marked increase in NLR emission from S1 to S3.  For the PCs, the fractional importance of each PC is shown. Notice that for S1 and S2, the new PC1 represents relationships found in ALL:PC2, namely the continuum slope and broad Balmer lines.  However, S3 is entirely dominated by NLR emission with the ALL:PC1 remaining as the S3:PC1.}
\label{finalcomp1a}
\end{figure*}

\begin{figure*}
\centering
\includegraphics[scale=0.9]{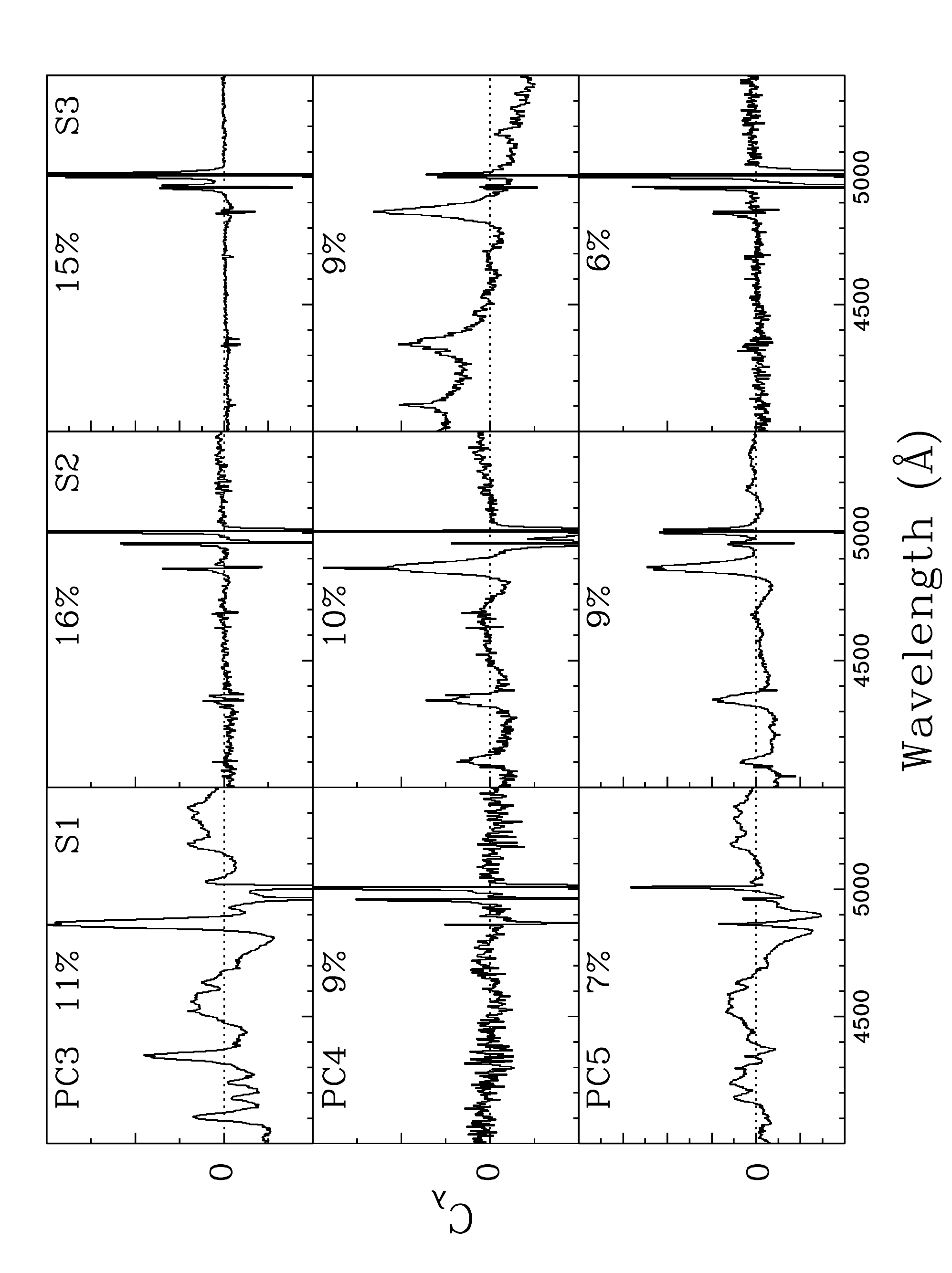}
\caption{New PC3, PC4, and PC5 for all three subsets, with columns as in Figure ~\ref{finalcomp1a}.  Note that S3 really is entirely dominated by NLR emission in all PCs except the S3:PC4, where the slope and Balmer lines show up again.  Also, S1:PC3 regains the traditional EV1 relationships, including the distinct anticorrelation between [\ion{O}{3}] and \ion{Fe}{2}, and the correlation with H$\beta$ linewidth.  For all subsets, PC4 and on each represents 10\% or less of the variance among objects and are therefore less physically significant.}
\label{finalcomp1b}
\end{figure*}

\subsubsection{Subset Results}

\begin{figure*}[!tp]
\centering
\subfigure{
\includegraphics[scale=0.35]{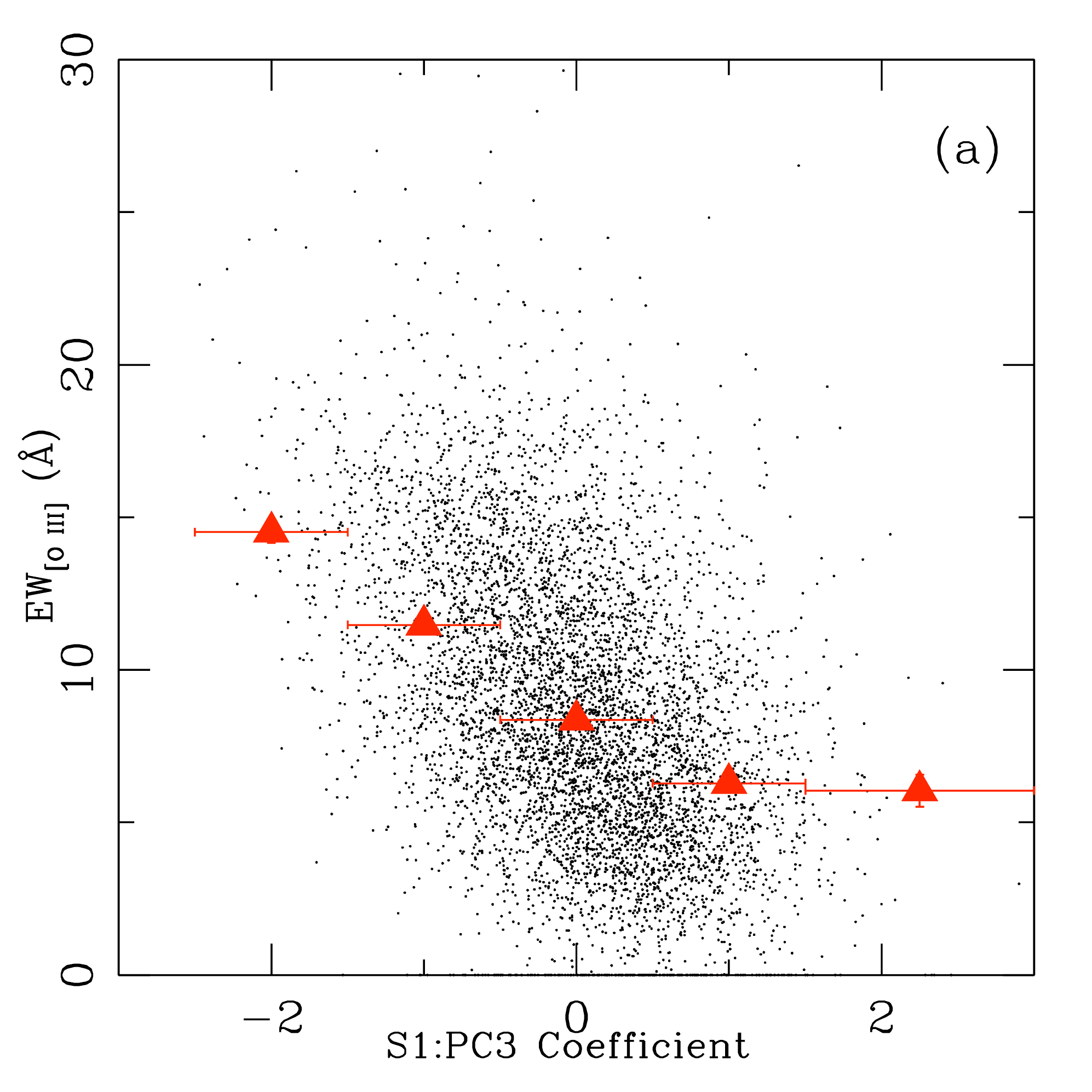}
\label{o3s1pc3}
}
\subfigure{
\includegraphics[scale=0.35]{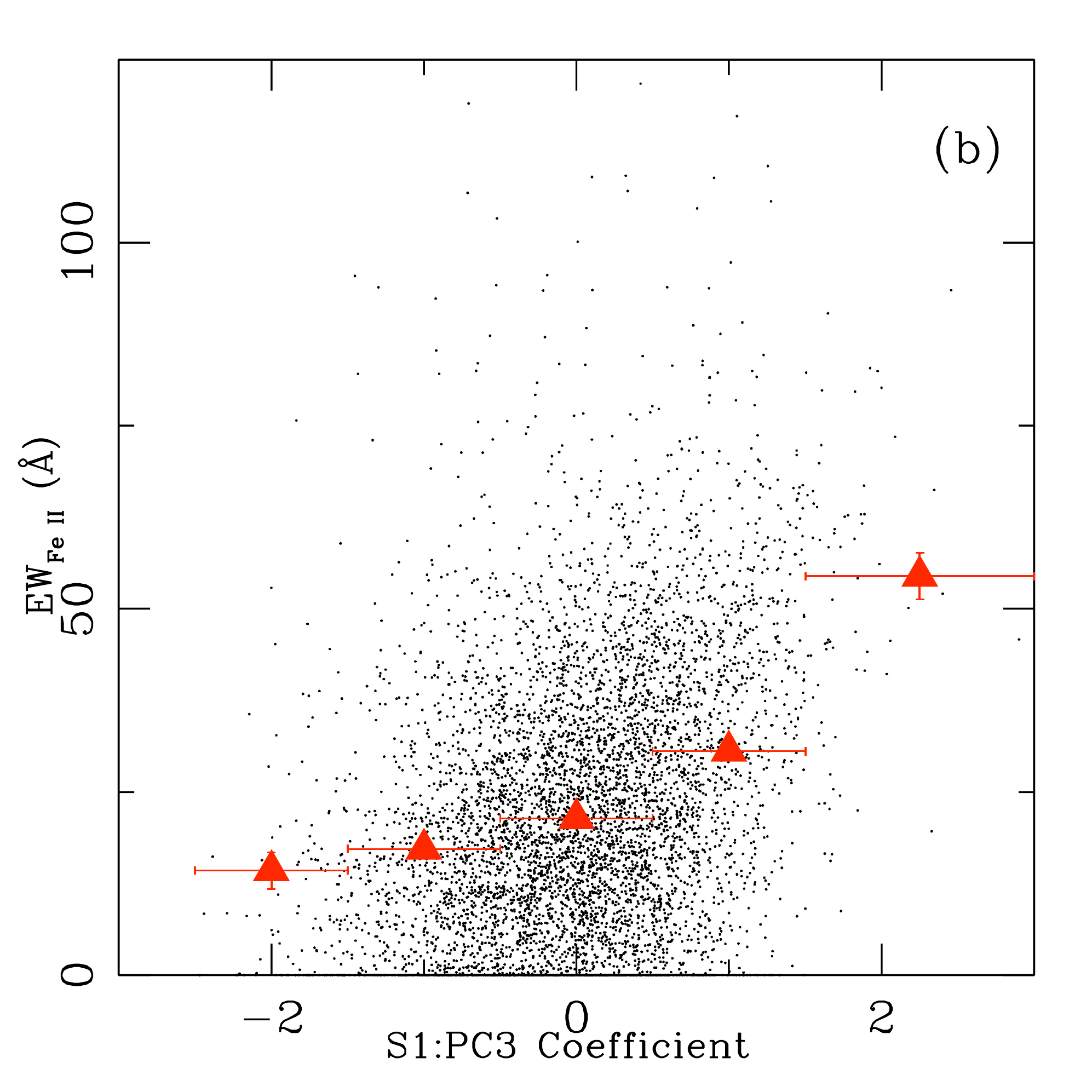}
\label{fe2s1pc3}
}
\subfigure{
\includegraphics[scale=0.35]{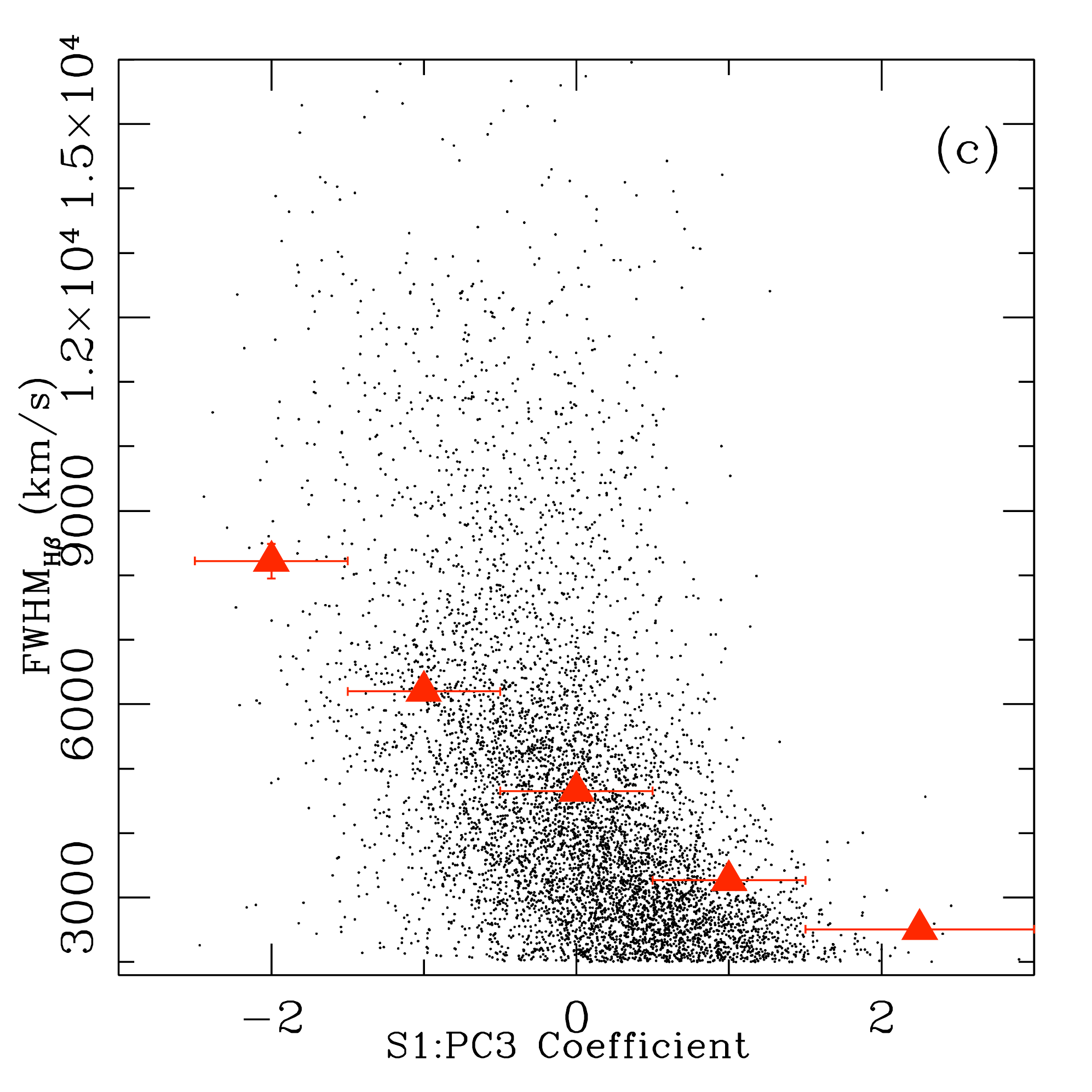}
\label{hbws1pc3}
}
\subfigure{
\includegraphics[scale=0.35]{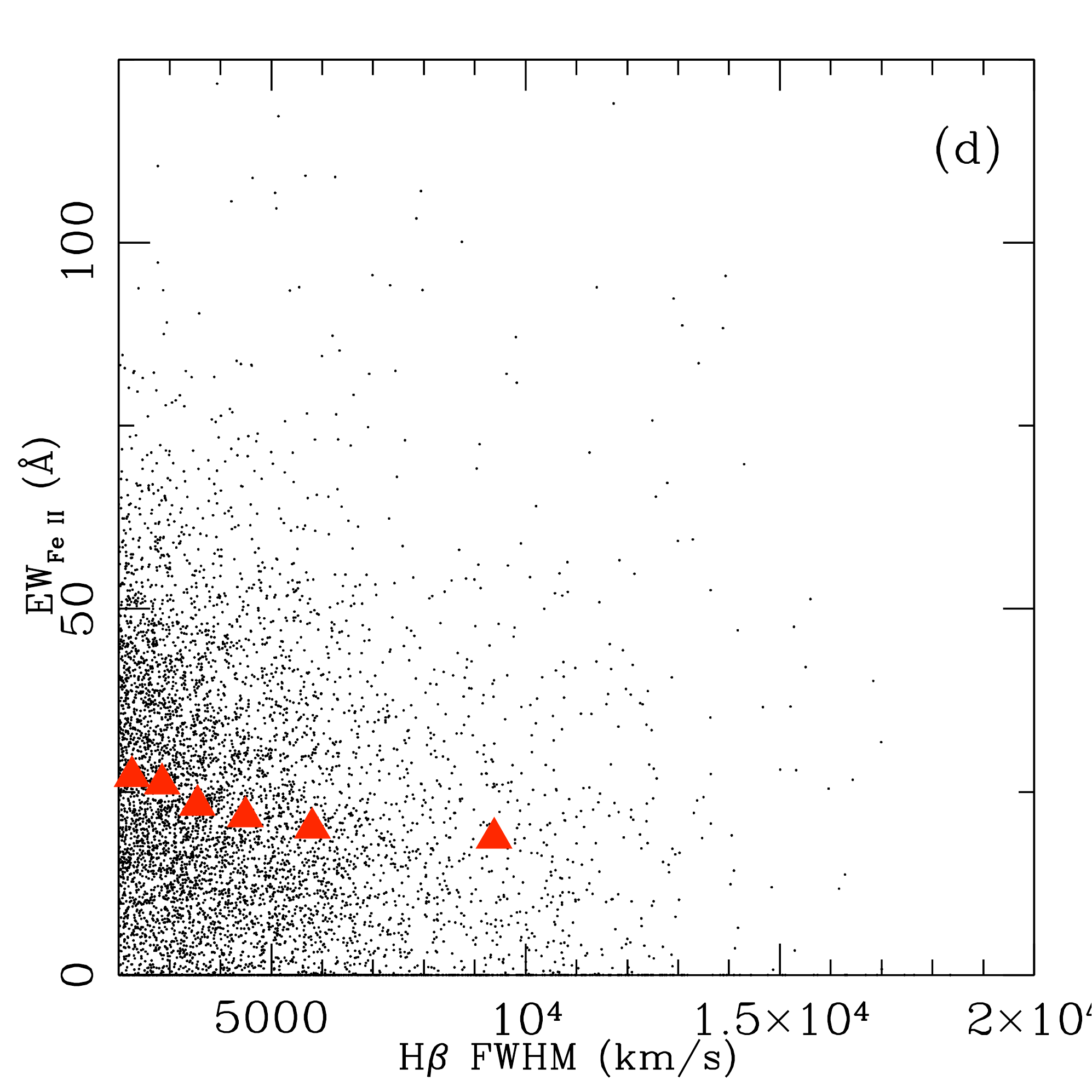}
\label{s1fe2hbw}
}
\caption{\textit{(a)} EW$_{\mathrm{[O III]}}$ versus S1:PC3 Coefficients.  The red triangles represent the mean of the data binned in the x-axis, so that each bin included at least 45 points, where error bars are the bin widths and standard deviations divided by $\sqrt{n-1}$ for each bin.  The error bars are sometimes smaller than the symbols.  
\textit{(b)} EW$_{\mathrm{Fe II}}$ versus S1:PC3 Coefficients.  Symbols are as in \textit{(a)}.
\textit{(c)} FWHM$_{\mathrm{H}\beta}$ versus S1:PC3 Coefficients.  Symbols are as in \textit{(a)}.
\textit{(d)} shows EW$_{\mathrm{Fe II}}$ versus FWHM$_{\mathrm{H}\beta}$ for S1 binned in FWHM$_{\mathrm{H}\beta}$ where error bars are standard deviations divided by $\sqrt{n-1}$. 
\label{fig:params1pc3}}
\end{figure*}

The resulting PCs for S1, which includes the low EW$_{\mathrm{[O III]}}$ objects where ALL:PC1 $< 0$, are in the left column of Figures ~\ref{finalcomp1a} and ~\ref{finalcomp1b}.  It includes $\sim$~70\% of the original data set and therefore illuminates the relationships among the bulk of broad-line AGN.  This subset consists of objects with a negative value of ALL:PC1, which means that reconstructing their spectra requires subtracting NLR emission from the original mean spectrum, so this subset includes the weakest EW$_{\mathrm{[O III]}}$.  As objects representing the strongest EW$_{\rm{[OIII]}}$ have been removed, ALL:PC2, representing the relationship between BLR emission and the continuum slope, becomes the first principal component (S1:PC1).  S1:PC2 includes a correlation among EWs of all  emission (Balmer lines and He II) and [\ion{O}{3}].  This subset also more closely recovers EV1 relationships in S1:PC3.  Note that there is a  ``W'' linewidth signature for H$\beta$ in S1:PC3, so this component links FWHM$_{\rm{H}\beta}$, \ion{Fe}{2} and [\ion{O}{3}] in a way similar to EV1.  To cross-check our qualitative interpretation of the PCs, we compare these three quantities with the coefficient of S1:PC3 (Figure ~\ref{fig:params1pc3}).  The EW$_{\mathrm{[O III]}}$ and FWHM$_{H\beta}$ anticorrelate with the S1:PC3 coefficient, with Spearman correlation coefficients of $-0.447$ and $-0.549$ respectively.  The EW$_{\mathrm{Fe II}}$ shows a correlation with S1:PC3 and a Spearman correlation coefficient of $0.310$.  These are the sense of relationships we would expect to see within BG92's EV1, and when we examined EW$_{\mathrm{Fe II}}$ versus FWHM$_{\mathrm{H}\beta}$ (Figure ~\ref{s1fe2hbw}), we found an inverse correlation for S1 (Spearman correlation coefficient of $-0.221$) although there is an enormous amount of scatter in the data.  Additional PCs account for only a few percent of the object-to-object variation within the subset, and so we do not interpret them.

The third subset, S3, which includes less than 5\% of the entire data set, is responsible for the most prominent relationships in ALL:PC1.  S3:PC1 (right column of Figures ~\ref{finalcomp1a}) shares the relationships seen in ALL:PC1.  Effectively, the extreme variation of EW$_{\mathrm{[O III]}}$ embodied in S3 compared to the rest of the entire sample was dominating the relationships that account for the most object-to-object variance among the 9046 object data set.  

As is evident after examining the S3:PCs, they are entirely dominated by variations in NLR emission lines, except for S3:PC4, which shows the same continuum slope and broad Balmer lines as ALL:PC2.  The S3:PCs represent correlations among NLR lines in terms of EW, asymmetry or shift, linewidth, and even stronger asymmetry in S3:PC5.  However, note that even though these 422 objects are clearly dominated by their NLR emission, they are selected to be broad-line AGN, not Type II AGN (see \S ~\ref{sec:bg92}).  The mean spectrum for this subset shows a distinct broad H$\beta$ contribution to the spectrum.  We will address in Section \S ~\ref{sec:outliers} possible reasons for these objects' extraordinarily strong NLR emission lines.

In the middle column of Figures ~\ref{finalcomp1a} and ~\ref{finalcomp1b}, we present results for the second subset, S2, which includes objects with ALL:PC1 coefficients between 0 and 7.  This subset, which is comprised of 2307 objects or about 1/4 of the entire data set, has properties intermediate between S1 and S3.  In S2:PC1, we basically recover ALL:PC2, which is a component dominated by broad Balmer lines and continuum shape.  S2:PC2 seems similar to ALL:PC1, dominated by NLR emission, but with additional continuum information as well.  S2:PC3 represents asymmetry in the NLR lines that could be caused by errors in the redshifts or real shifts in the emission lines.

\section{Comparison with Previous Work}
\label{sec:bg92}

One reason for our investigation was to examine the EV1 relationships among a larger, more diverse sample than BG92 and S03.  While S03 did recover the traditional EV1 relationships, including the anticorrelation of [\ion{O}{3}] and Fe II, and the simultaneous correlation of Fe II and FWHM$_{H\beta}$, the PCs from our original entire data set do not show these relationships clearly at all.  This implies that the crucial difference between our results and theirs have to do with the samples themselves.

Because ALL:PC1 is dominated by NLR emission, where BG92 and S03 did not find a similar component with such dramatic NLR characteristics, we wondered if the main difference between our SDSS sample and theirs was the NLR properties of the samples.  To investigate this, we compared the distribution of EW$_{\mathrm{[O III]}}$ from BG92 with our distribution (see the histogram in Figure ~\ref{o3allpc1}).  Our range of EW$_{\mathrm{[O III]}}$ extended to objects with three times stronger [\ion{O}{3}].  A Kolmogorov-Smirnov (K-S) test, which returns the maximum discrepancies between the two samples using both one-sided cases and the two-sided case, results in an inconclusive P=0.04 likelihood that the two EW$_{\mathrm{[O III]}}$ distributions are drawn from the same parent distribution.  In Figure ~\ref{pcvpc}, we include the PG quasars that have spectroscopic observations from the SDSS.  There are only two PG quasars that fall in our strong [\ion{O}{3}] subset, S3, but there are few enough points we cannot conclude that the [\ion{O}{3}] distribution is significantly different.  It is interesting to note that all the PG quasars have positive ALL:PC2, indicating that they fall among the luminous, blue objects within ALL:PC2-space.  

\begin{figure}[!tp]
\centering
\includegraphics[scale=0.5]{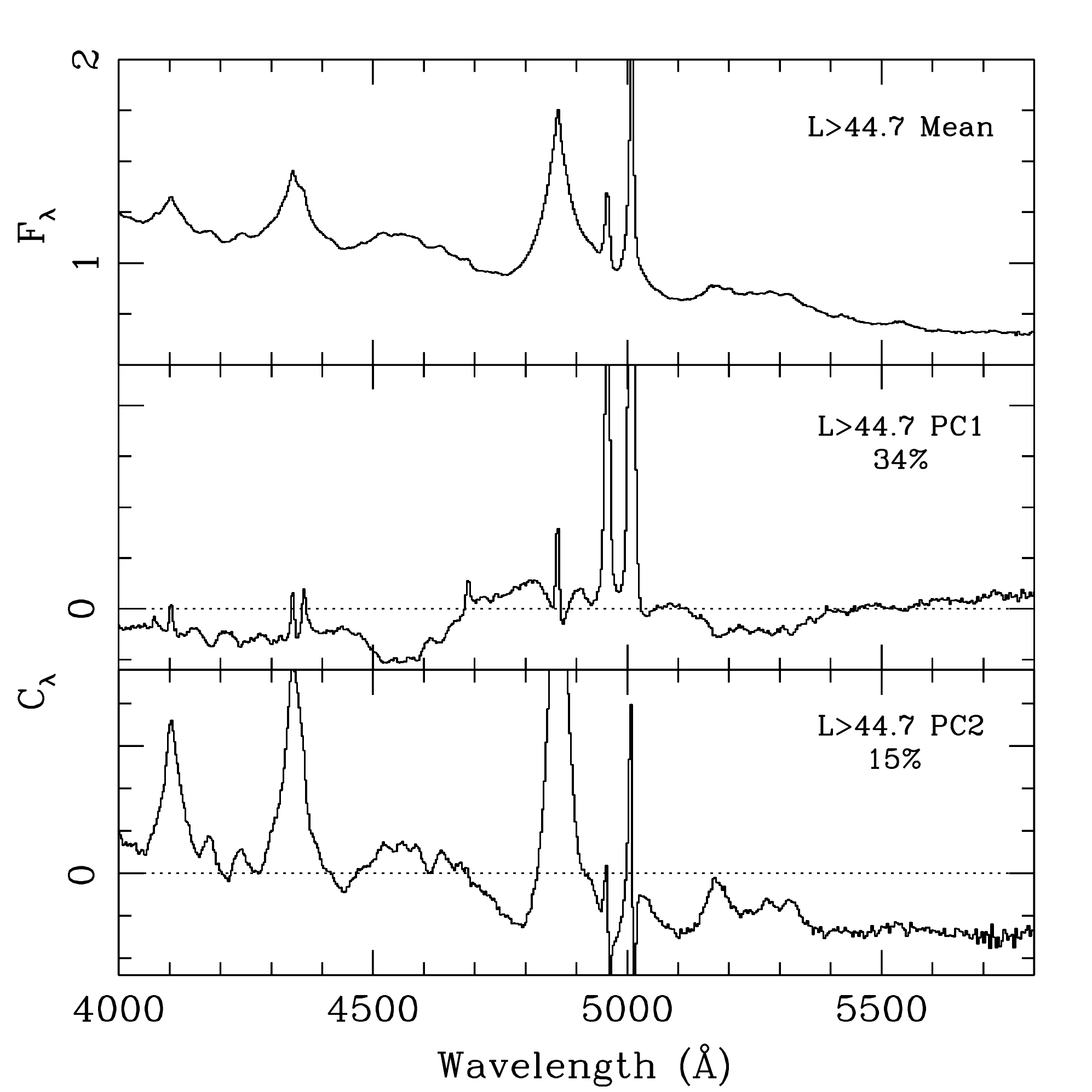}
\caption{Mean spectrum, PC1, and PC2 for the the 584 objects with $L_{5100 \mbox{\AA}} > 10^{44.7}$ erg s$^{-1}$.  PC1 recovers the BG92 \ion{Fe}{2} - [\ion{O}{3}] anticorrelation and H$\beta$ linewidth dependence and PC2 incorporates the continuum slope and the broad Balmer lines.}
\label{w_Lge447comp1}
\end{figure}

\begin{figure*}[!tp]
\centering
\subfigure{
\includegraphics[scale=0.35]{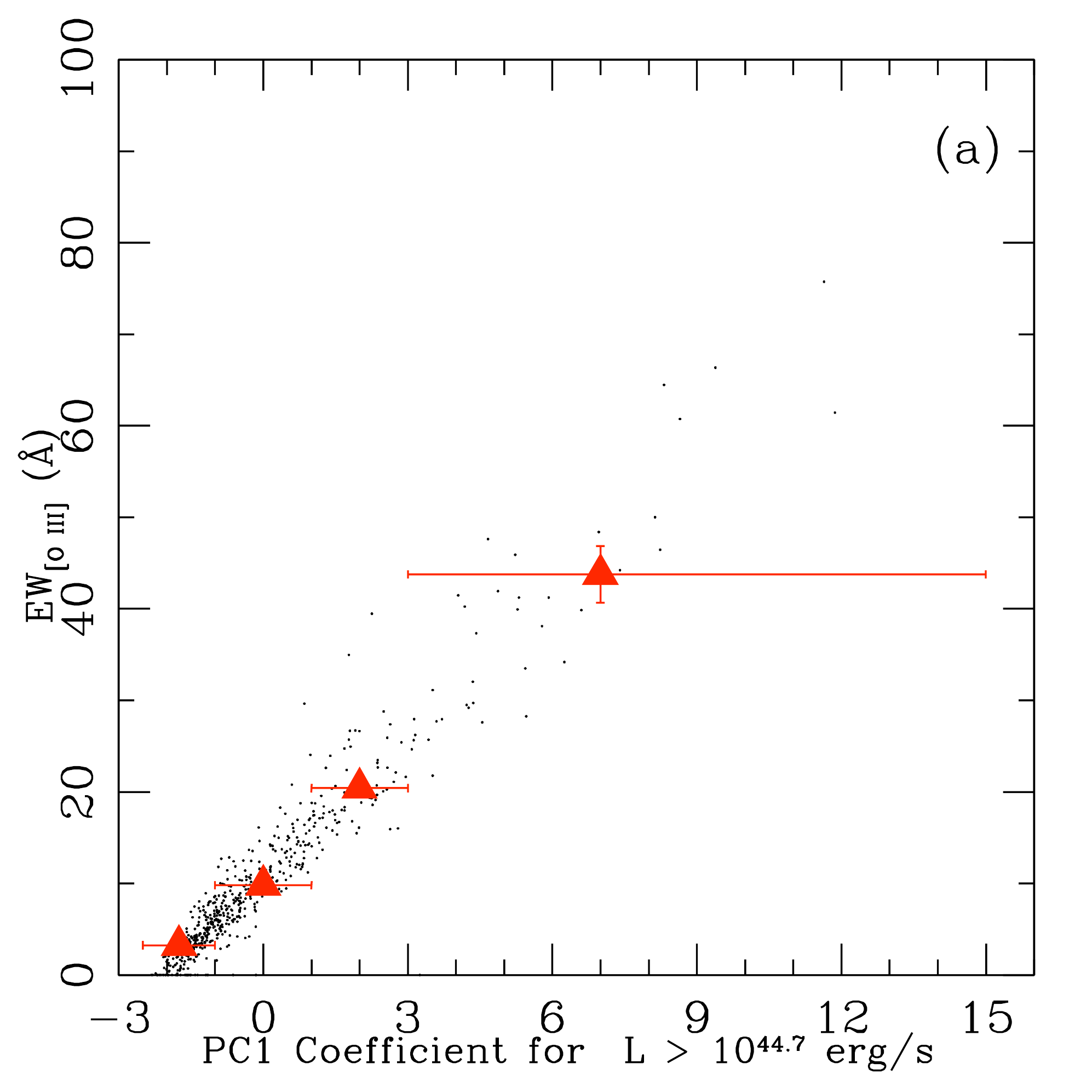}
\label{o3Lpc1}
}
\subfigure{
\includegraphics[scale=0.35]{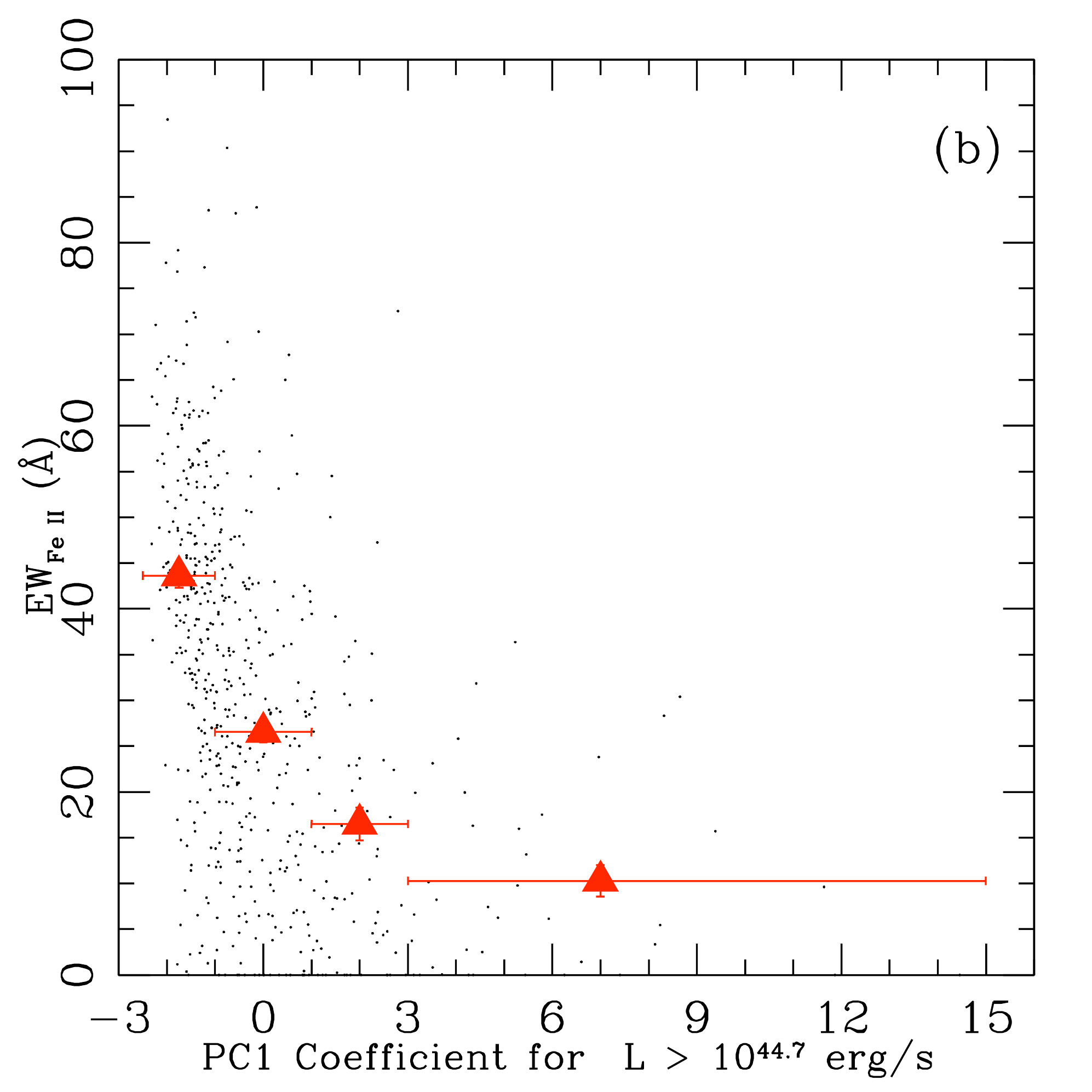}
\label{fe2Lpc1}
}
\subfigure{
\includegraphics[scale=0.35]{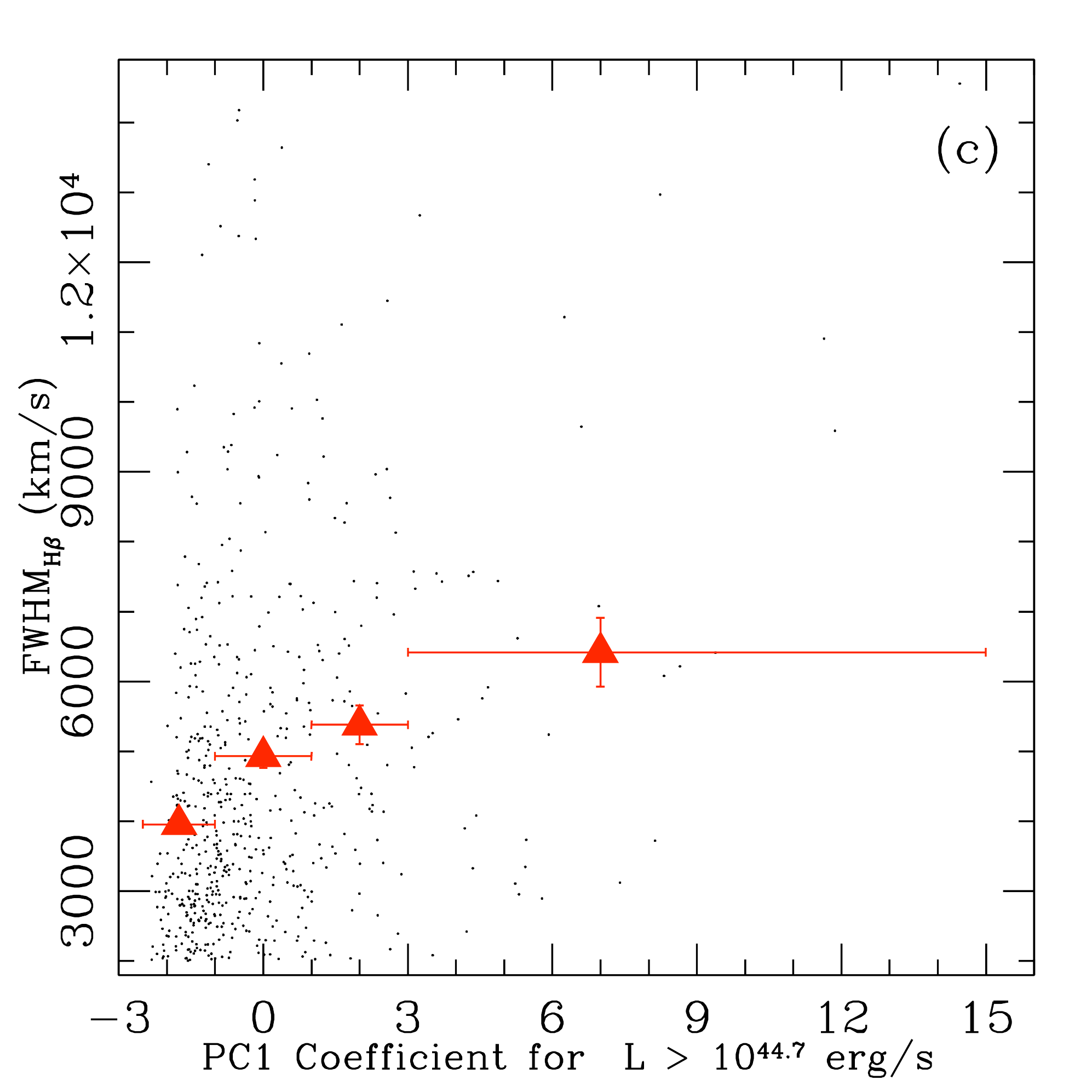}
\label{hbwLpc1}
}
\subfigure{
\includegraphics[scale=0.35]{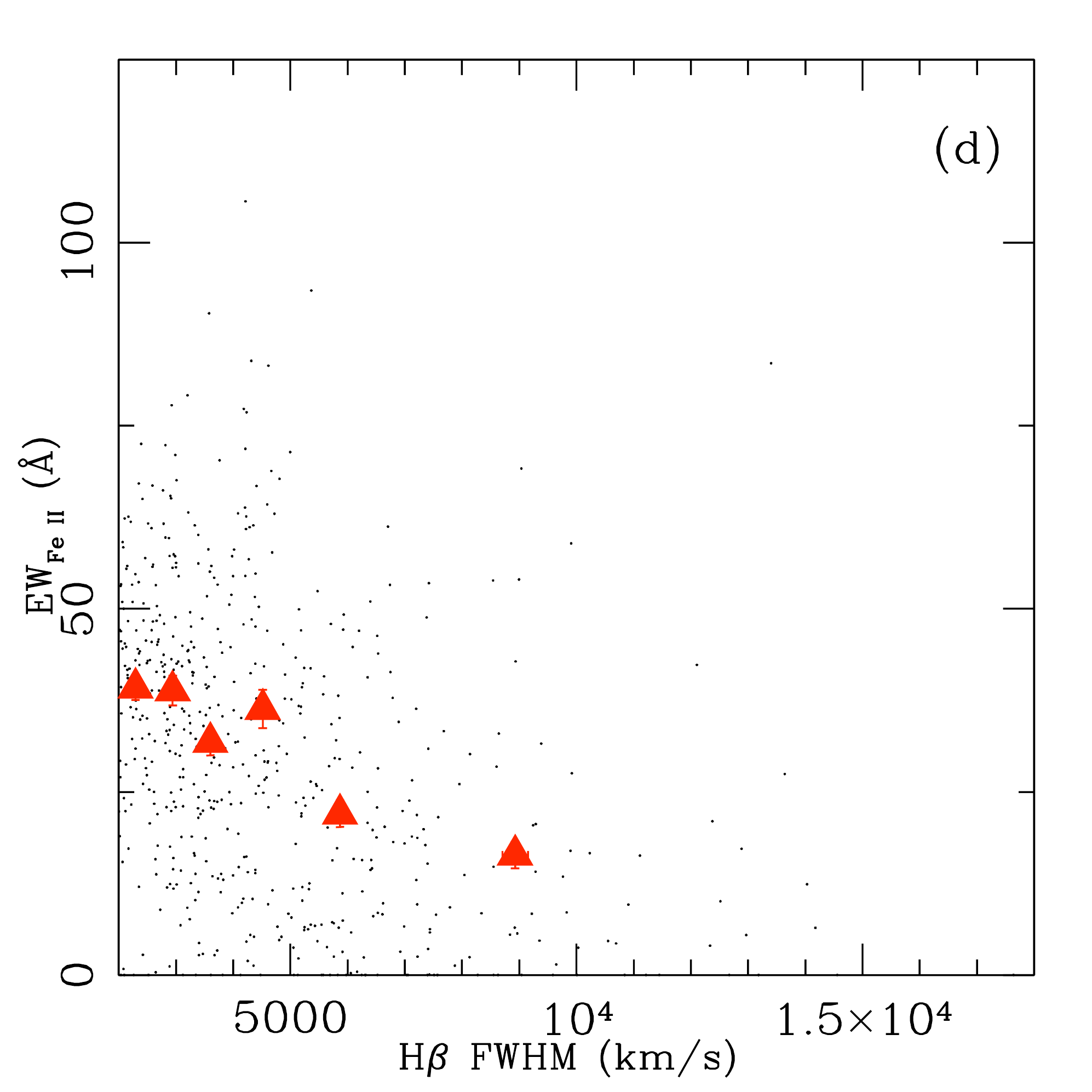}
\label{Lfe2hbw}
}
\caption{\textit{(a)} EW$_{\mathrm{[O III]}}$ versus PC1 Coefficients for the sample with $L_{5100 \mbox{\AA}} > 10^{44.7}$ erg s$^{-1}$.  The red triangles represent the mean of the data binned in the x-axis, so that each bin included at least 39 points, where error bars are the bin widths and standard deviations divided by $\sqrt{n-1}$ for each bin.  The error bars are sometimes smaller than the symbols.  
\textit{(b)} EW$_{\mathrm{Fe II}}$ versus PC1 Coefficients for the sample with $L_{5100 \mbox{\AA}} > 10^{44.7}$ erg s$^{-1}$.  Symbols are as in \textit{(a)}.
\textit{(c)} FWHM$_{\mathrm{H}\beta}$ versus PC1 Coefficients for the sample with $L_{5100 \mbox{\AA}} > 10^{44.7}$ erg s$^{-1}$.  Symbols are as in \textit{(a)}.
\textit{(d)} shows EW$_{\mathrm{Fe II}}$ versus FWHM$_{\mathrm{H}\beta}$ for the high luminosity sample binned in FWHM$_{\mathrm{H}\beta}$ where error bars are standard deviations divided by $\sqrt{n-1}$. 
\label{fig:paramLpc1}}
\end{figure*}

There is a much more crucial difference between the two subsamples, namely luminosity.  The PG quasars used by BG92 and S03 include only high luminosity sources with a nominal absolute magnitude of M$_{\rm{V}}\le -23$ ($L_{5100 \mbox{\AA}} \gtrsim 10^{44.7}$ erg s$^{-1}$) although $\sim$20\% of the PG quasars are below this nominal limit.  This is a striking difference from our luminosity distribution, where only $\sim$6\% of the objects have $L_{5100 \mbox{\AA}} > 10^{44.7}$ erg s$^{-1}$.  Indeed the EW$_{\mathrm{[O III]}}$ distribution for our high-$L_{5100 \mbox{\AA}}$ objects is quite similar to that of BG92, although their inclusion of fainter objects makes an exact comparison difficult.  When we performed SPCA on the 584 objects with $L_{5100 \mbox{\AA}} \gtrsim 10^{44.7}$ erg s$^{-1}$, we did in fact recover a PC1 that is similar to S03's first principal component, although ours still has more NLR contribution.  It has a more striking [\ion{O}{3}] - \ion{Fe}{2} anticorrelation than ALL:PC1 and includes H$\beta$ linewidth dependence (Figure ~\ref{w_Lge447comp1}).  Again, these relationships are confirmed by the measurements shown in Figure ~\ref{fig:paramLpc1}.  In fact, plotting EW$_{\mathrm{Fe II}}$ versus FWHM$_{\mathrm{H}\beta}$ (Figure ~\ref{Lfe2hbw}) for these objects shows a dramatic correlation between the two (Spearman correlation coefficient of $-0.411$).  This confirms that we are recovering the correlations from BG92's EV1 in this set of high $L$ objects, although there is a large amount of scatter.

\begin{figure*}[!tp]
\centering
\subfigure{
\includegraphics[scale=0.35]{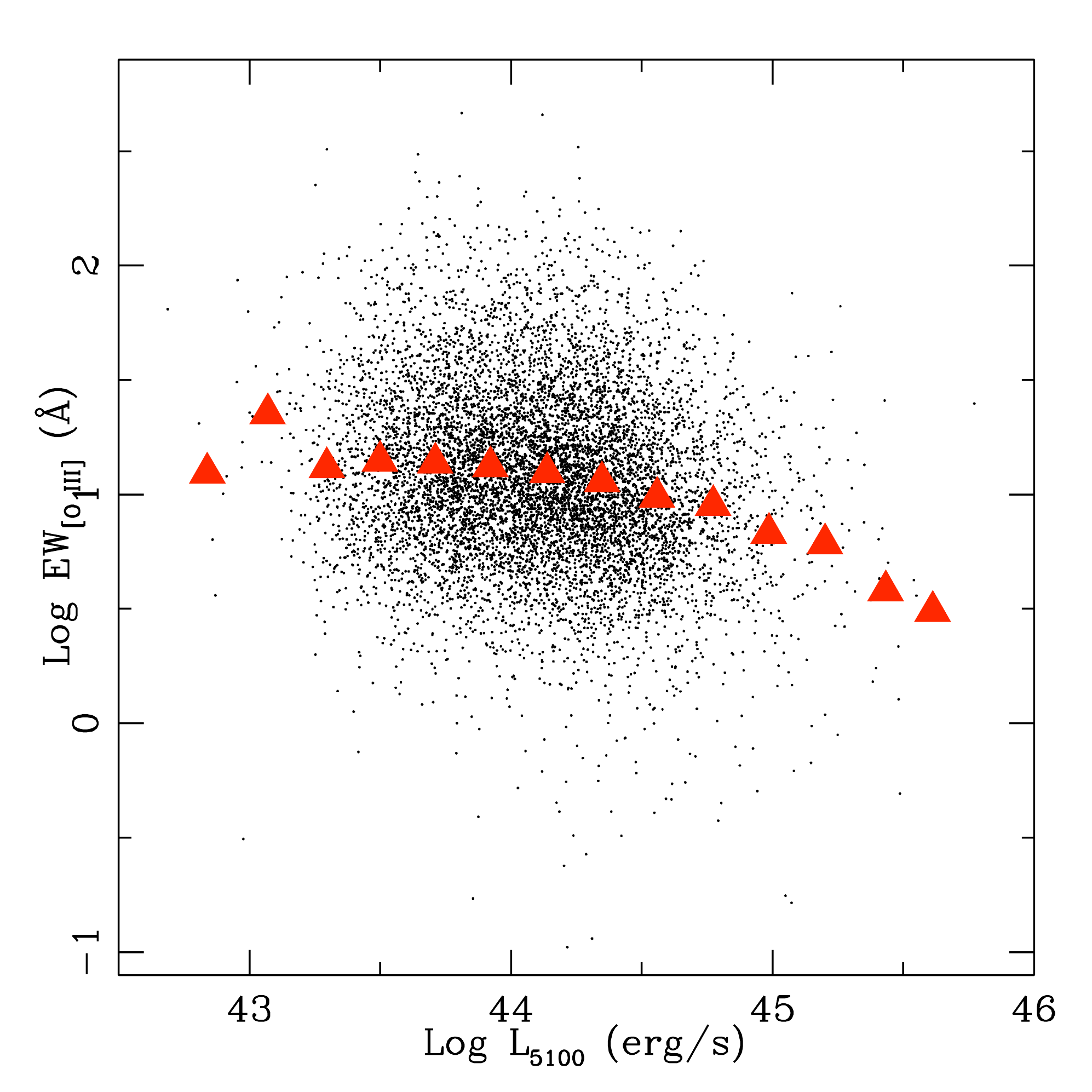}
}
\subfigure{
\includegraphics[scale=0.35]{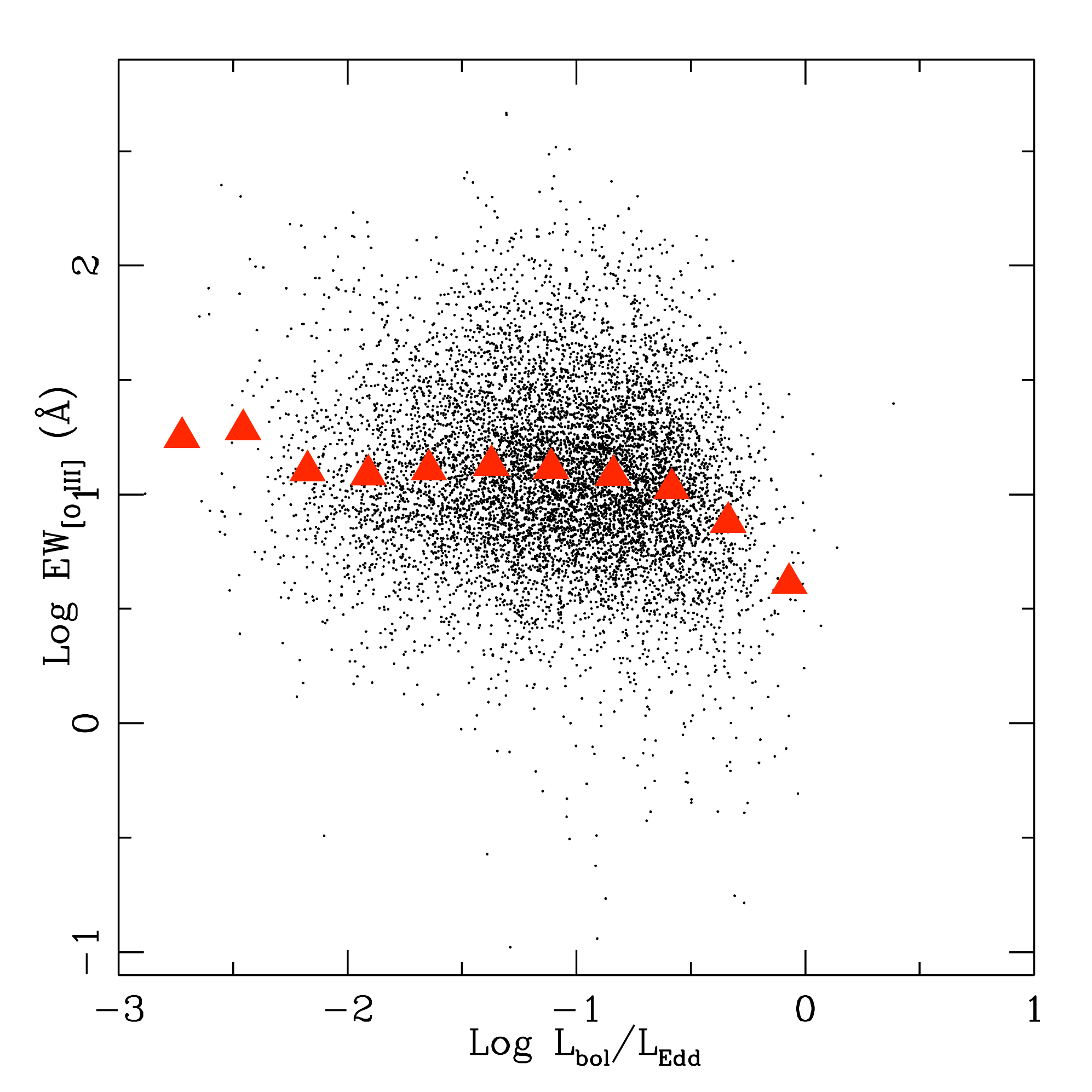}
}
\caption{The logarithm of EW$_{\mathrm{[O III]}}$ versus Log $L_{5100 \mbox{\AA}}$ and $L_{\rm{bol}}/L_{\rm{Edd}}$ for the entire sample.  From our data we are unable to disentangle whether the drop in EW$_{\mathrm{[O III]}}$ at high $L$ is strictly dependent on luminosity or Eddington ratio, but for low $L$ and $L_{\rm{bol}}/L_{\rm{Edd}}$ there is no correlation.}
\label{o3ewvs}
\end{figure*}

The luminosity difference between the PG quasars and our original sample apparently leads to the different NLR emission in ALL:PC1.  However, within our SDSS sample, we found no conclusive difference between the $L_{5100 \mbox{\AA}}$ distributions of the whole SDSS sample and S3, where the K-S test results in 0.36 probability that they are from the same parent distribution (see \S ~\ref{sec:spec}).  Additionally, at low $L_{5100 \mbox{\AA}}$, we do not see a clear correlation between EW$_{\mathrm{[O III]}}$ and $L_{5100 \mbox{\AA}}$ (see Fig. ~\ref{o3ewvs}), possibly due to the large scatter in the data, but clearly there is a drop in EW$_{\mathrm{[O III]}}$ at very high $L_{5100 \mbox{\AA}}$.  \cite{2002MNRAS.337..275C} also did not find a decrease in EW$_{\mathrm{[O III]}}$ as a function of $L$ for a sample spanning a similar range of $L$ and $z$ as this one, although they do find a trend with other NLR emission lines.  However, they explain that the effect is
not seen in [\ion{O}{3}] because it is observed at lower $z$, and in the local universe there is a lack of especially high-$L$ objects.  Other studies with targets at higher $L$, such as  \cite{2006A&A...453..525N, 2004ApJ...614..558N} and \cite{2009A&A...495...83M}, do find a clear anticorrelation between EW$_{\mathrm{[O III]}}$ and $L$.  It would
be interesting to know if this drop in EW$_{\mathrm{[O III]}}$ was strictly
dependent on $L$ or more dependent on $L_{\rm{bol}}/L_{\rm{Edd}}$, to better understand the physical relationships in AGN since, for example, opening angle is luminosity-dependent \citep[e.g.][]{1991MNRAS.252..586L, 2003NewAR..47..211S}.  It has been hypothesized \citep{2005ApJ...620..629D, 2005MNRAS.360..565S, 2006A&A...453..525N} that variations in opening angle, which depends on $L$, impact the strength of [\ion{O}{3}] emission that is observed.
However, the fact that narrow-line Seyfert 1 galaxies (NLS1s), which are thought to be in a particularly high
accretion state, are distinguished by extreme EV1 properties does
suggest that there is a correlation between
$L_{\rm{bol}}/L_{\rm{Edd}}$ and EW$_{\mathrm{[O III]}}$.  (Note that since we
have excluded NLS1s from our sample, we may have removed the highest
$L_{\rm{bol}}/L_{\rm{Edd}}$, low EW$_{\mathrm{[O III]}}$ objects from
our analysis.)  The Eddington ratio (Fig. ~\ref{o3ewvs}) of our sample
does not seem to have a correlation with the [\ion{O}{3}] strength, except at high Eddington fractions.  With
these data, we cannot distinguish whether this drop in EW$_{\mathrm{[O III]}}$ is an effect of $L$ or $L_{\rm{bol}}/L_{\rm{Edd}}$.

\section{The Strong [\ion{O}{3}] Subset}
\label{sec:outliers}

In this section we search for the cause of the large 
narrow-line equivalent widths in the spectra of the
AGN in the S3 sample.
We first look at the radio loudness of the S3 objects but find
no correlation.
We then look for signatures of extra obscuration of the AGN in S3 relative to the entire sample
due to orientation effects, which might lower the AGN continuum
and increase the narrow-line equivalent widths, but
find no strong signatures.
Finally we search for other differences between the spectral
properties of the AGN in the S3 sample and those in the S1 and S2 samples.
We conclude this section with a brief discussion suggesting that larger
covering factors account for the large equivalent widths.

\subsection{Radio Properties}

\begin{figure}[!tp]
\centering
\includegraphics[scale=0.45]{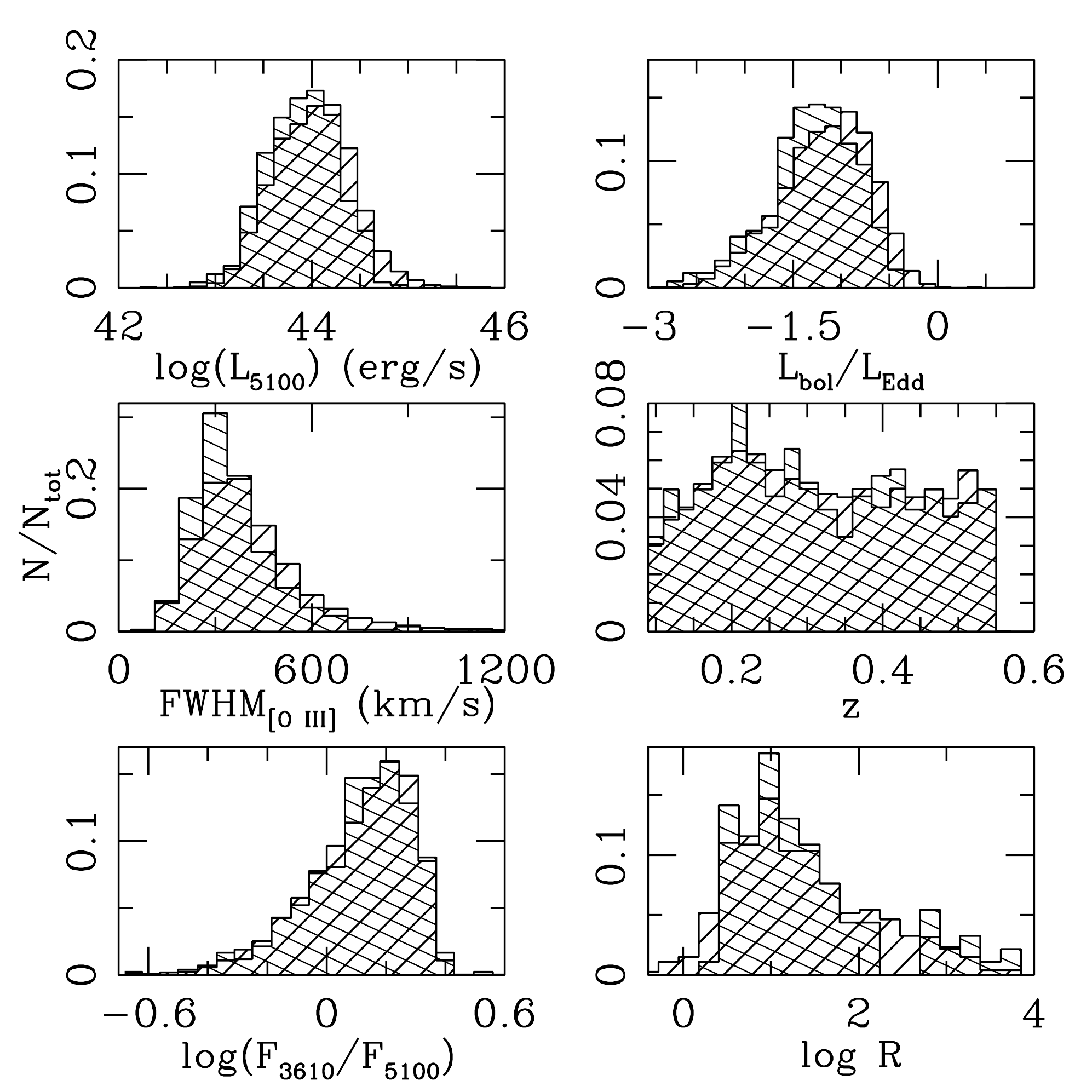}
\caption{Histograms for the distribution of the entire data set, with upward-right slanting hashes, and just S3, with downward-right slanting hashes, for luminosity, Eddington ratio, FWHM$_{\mathrm{[O III]}}$, continuum color, redshift, and EW$_{\mathrm{[O III]}}$.  The luminosity and redshift distributions are quite similar, but the objects in S3 have notably narrower and stronger [\ion{O}{3}] emission.}
\label{hist}
\end{figure}

\begin{figure}[!tp]
\centering
\includegraphics[scale=0.5]{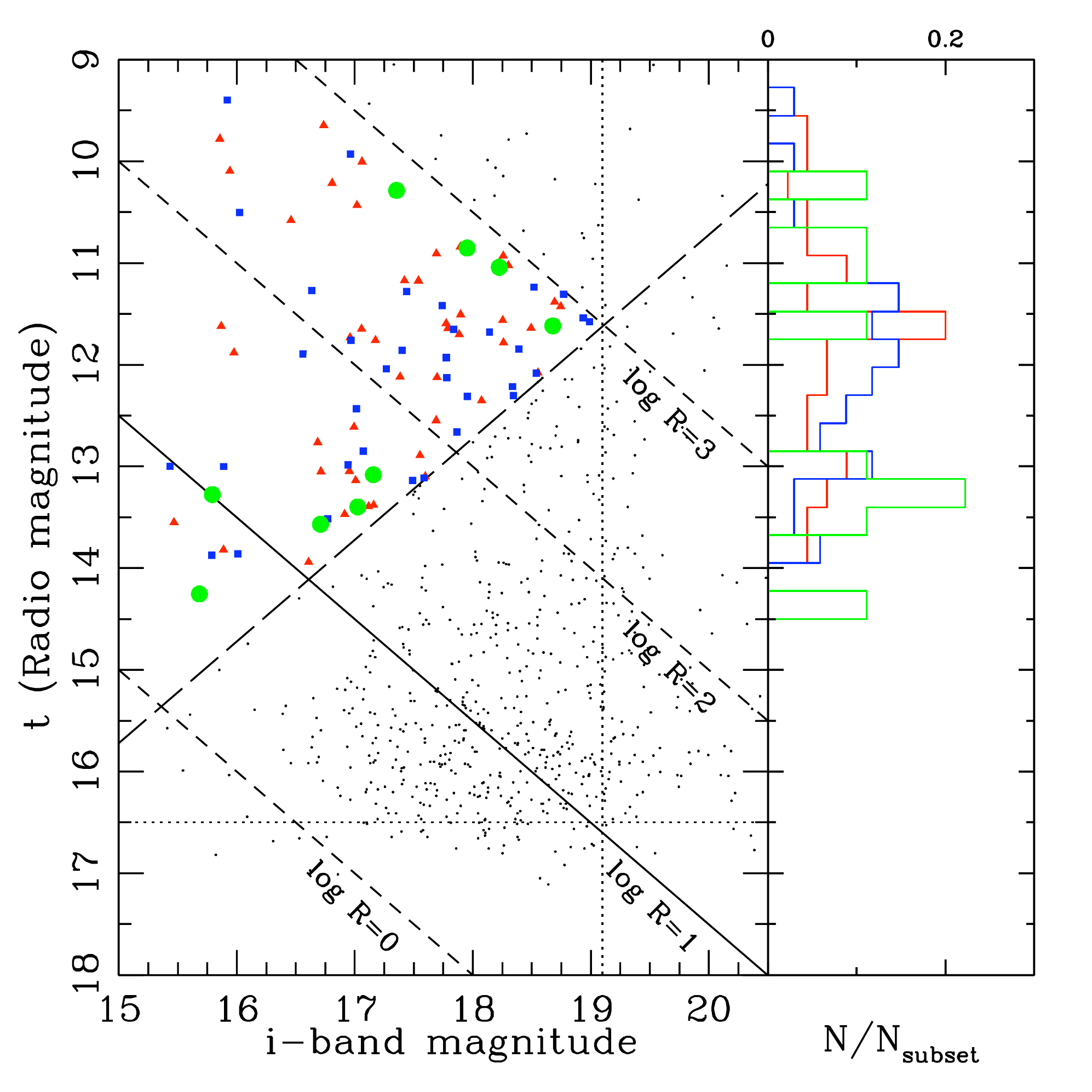}
\caption{Radio magnitude, $t=-2.5 \rm{ log}(F_{rad}/3631 Jy)$, vs. \textit{i}-band magnitude for the 735 objects with radio detections.  The dotted lines denote the magnitude limits of SDSS and FIRST, at $i < 19.1$ and $t < 16.5$ respectively.  The solid line is log $R=1$, demarcating the transition between radio-loud and radio-quiet objects. The dashed lines are lines of constant log $R$ equaling 0, 2, and 3 from bottom to top.  The long-dashed line, perpendicular to the lines of constant log $R$ and passing through the intersection of log $R = 3$ with the SDSS magnitude limit, denotes a complete sample.  Red triangles represent S1, blue squares represent S2, and green circles represent S3.  The colored histograms represent the stringent sample distribution in \textit{t}, with colors as before.}
\label{radio}
\end{figure}

Past authors \citep[e.g.][and references therein]{2001ApJ...555..650H} investigated the degree to which radio
power correlates with $L_{\mathrm{[O III]}}$, and found that for
radio-quiet sources, the correlation is quite strong.  We investigate
the radio properties of the objects in the strong [\ion{O}{3}] subset to see if
they have correspondingly strong radio emission.  We
cross-correlated our list of 9046 SDSS objects with the Faint Images
of the Radio Sky at Twenty-Centimeters (FIRST) catalog of radio
sources.  The FIRST survey was conducted using the Very Large Array
(VLA) and covers over 9000 square degrees of sky.  The detection
threshold is 1mJy at 1.4 GHz in the observed frame.  We find 735 radio
detections within 2" of the optical source for our sample.  We chose a
2" search radius as it has been shown that the number of meaningful
correlations with optical AGN detections drops off rapidly outside 2"
\citep{2002AJ....124.2364I,2009AJ....137...42R,2008A&A...488..145H}.

Of the 9046 objects in our entire data set, 8460 lie within the FIRST
footprint, rejecting less than 10\%.  Therefore, the 735 detections
represent 8\% of the data set within the observing area of FIRST.
However, of the 422 objects in S3, 404 were within the footprint, with
92 detections, yielding a fraction of 22\% detected.  Thus, a higher
proportion of the strong [\ion{O}{3}] objects have detectable radio
emission.

The S3 objects have the same redshift distribution as the entire data
set (Figure ~\ref{hist}), so the higher detection percentage for S3 is
possibly meaningful but susceptible to various selection biases.  As
an extra precaution, we follow the prescription in
\cite{2002AJ....124.2364I} for defining a ``stringent radio sample'' that
is not subject to a Malmquist bias for detections from either
FIRST or SDSS.  The Malmquist bias results in extra detections near
the flux limit of a given survey due to statistical errors in
detection.

We selected a complete sample above,
well within the flux limits of both SDSS and FIRST. In Figure
~\ref{radio}, we have plotted the observed radio magnitude, $t=-2.5
\rm{ log}(F_{rad}/3631 Jy)$, where we use the integrated radio flux
density of the detected objects versus their $i$-band magnitude
following \cite{2002AJ....124.2364I}. We calculate the radio to
optical flux density ratio as log $R=0.4(i-t)$ \citep{2002AJ....124.2364I}.  The flux limits of both
SDSS, at $i < 19.1$, and FIRST, at $t < 16.5$, are shown as the dotted
vertical and horizontal lines.  To simplify the analysis, we did not
include a K-correction, but chose to use the observed magnitudes for
the radio and \textit{i}-band.

To select a complete sample, we wanted to span the full range observed
in log $R$, where the short-dashed lines are lines of constant log
$R$.  We selected the stringent radio sample by taking objects that
lie above the long-dashed line perpendicular to the lines of constant
log $R$ and within $0 <$log~$R < 3$, which then includes only objects
safely within the flux limits.  The solid line denotes log $R = 1$,
which separates radio-loud objects (log $R > 1$) from radio-quiet
objects (log $R < 1$).  Within this stringent sample, we have plotted
objects included in S1 as red triangles, objects in S2 as blue
squares, and objects in S3 as green circles.

If radio properties were really physically related to extreme NLR
emission, then we would expect the S3 objects
within the stringent radio sample to have significantly higher values
of log $R$ than those in S1 and S2.  However, within the stringent
sample, the distribution of S3 objects is not noticeably biased toward
the radio-loud side of the sample.  Figure ~\ref{hist} includes the
histogram of the distribution of S3 objects in $t$, compared with
the rest of the stringent sample.  While the K-S test results in only
a 0.6\% chance that the whole stringent radio sample and S3 objects
within it are from the same parent distribution, there does not appear
to be a striking correlation of radio-loudness with [\ion{O}{3}]
emission.  One caveat to this conclusion is the small number
statistics within the stringent radio sample, as there are only 9
objects from S3.  The radio properties might be somewhat different
between the S3 objects and the rest of the stringent sample, and S3
does appear to have a higher percentage of objects that are detected
in the radio compared with the entire data set, but there is not a
simple increase of radio-loudness with EW$_{\mathrm{[O
III]}}$.

\subsection{``Buried AGN'' Hypothesis}

In the orientation Unified Scheme, AGN are surrounded by 
an optically-thick dusty torus \citep[e.g.][]{1993ARA&A..31..473A}.  
If the observer's 
line of sight to the nucleus misses the torus, 
the observer sees both the broad-line and narrow-line regions; but if the 
line of sight passes through the torus, the BLR is obscured 
-- the AGN is ``buried''--  and the observer sees only a narrow-line 
spectrum from unobscured gas far from nucleus.
Indeed, given the proper orientation, an optically-thick cloud
of dust anywhere in the host galaxy can obscure the AGN.
If, for instance, the line of sight to the AGN just grazes the torus, the
AGN is partially obscured.
The broad-line emission and the AGN continuum appear weaker,
the latter leading
to higher equivalent widths for the narrow lines.
Examples of these objects include Seyfert 1.5-1.8
galaxies \citep{1993ARA&A..31..473A,1995ApJ...440..578T,1997ApJ...489L.137L,1999AJ....118.1963C} that have strong
narrow-line emission and weak broad-line emission.

Our composite S3 spectrum closely resembles the spectra of Seyfert
1.5-1.8 galaxies.  Does obscuration of objects in S3 cause them to have high EW$_{\mathrm{[O III]}}$?  
We first note that the S3 sample definitely includes buried AGN,
possibly many buried AGN.
We will see that the broad-line H$\alpha$/H$\beta$ flux ratio is 
large in some of the S3 AGN, a clear signal of reddening; 
a visual inspection of individual spectra in the S3 sample quickly 
finds classic Seyfert 1.5-1.8 objects; and
component S3:PC4 is a broad-line, blue-continuum
component devoid of NLR features that is introduced by variable
obscuration and reddening.
The question, though, is whether the obscuration in the S3 sample is causing higher EW$_{\mathrm{[O III]}}$.  We first investigate whether 
the obscuration in the S3 sample as a whole is different from the obscuration in
the S1 and S2 samples. 
There are several tests that might answer this question.

First, if the AGN in the S3 sample are more obscured than
the AGN in the S1 and S2 samples, they might be fainter
than the AGN in S1 and S2.
The mean $L_{5100 \mbox{\AA}}$ of each subset is $1.39 \times
10^{44}$, $1.06 \times 10^{44}$, and $9.68 \times 10^{43}$ erg s$^{-1}$for S1, S2,
and S3 respectively.
Thus, the mean $L_{5100 \mbox{\AA}}$ of the S3 AGN is, indeed, lower than the mean 
luminosity of the AGN in the other sets, but the maximum difference
is a factor of just 1.4, much less than the difference in mean
[\ion{O}{3}] equivalent widths.  This test is not a strong one, given the steep luminosity function of AGN.  
We also  compared the distribution of $L_{5100 \mbox{\AA}}$ luminosities within
the S3 sample to the distribution for the entire sample 
(see Figure~\ref{hist}).
(The S3 objects comprise a small fraction of the entire 
sample, so a comparison of the S3 sample with the entire sample is essentially
a comparison with S3 to S1 plus S2.)
A K-S test yields a $P=0.36$ probability that the S3 sample comes from the same 
parent distribution as the entire sample.
Thus there is no evidence that the luminosities of the S3 AGN
are lower than the luminosities of the entire AGN sample by
enough to account for the higher narrow-line EWs.

Second, if the AGN in the S3 sample are more obscured than
the entire sample, they should be more reddened and
their colors should be different.
We defined the color to be the log of the ratio of the median fluxes
in 20 \AA{} bins at 3610 \AA{} and 5100 \AA{} that are
relatively free of emission lines.  
We compared the color
distribution for S3 to that of the entire data set (see Figure ~\ref{hist}).
A K-S test finds a $P=0.78$ probability that the two distributions are
drawn from the same parent sample.
Thus this test finds no evidence for more reddening in the S3 sample than
in the entire sample.

Third, because the continuum and the BLR emission lines do not
necessarily arise in the same spatial regions, they could be 
reddened differently.
Therefore we tested whether the behavior
of the H$\alpha$/H$\beta$ flux ratio, which is a good reddening 
indicator for the BLR region, is different from the behavior 
of the continuum colors.
We measured the Balmer decrement for the 51 objects
in the S3 sample with $z < 0.35$ so that both H$\alpha$ and H$\beta$
are visible in their SDSS spectra.
We fitted their spectra using the IRAF package \textit{specfit} between
6400--6800 \AA{}, using 8 components for the fits: a
fixed power law for the continuum, Gaussians for the [\ion{S}{2}] $\lambda\lambda$6717,
6730 doublet, the [\ion{N}{2}] $\lambda\lambda$6547, 6583 doublet,
the narrow component of H$\alpha$, and two Gaussians
for the broad component of H$\alpha$, with the narrow-lines all
forced to have the same width.  
The Balmer decrement for the BLR is the 
ratio of broad H$\alpha$ flux to broad H$\beta$ flux.  
We measured the mean Balmer decrement for the 51 objects as 8.311, and it ranged from 3.37 -- 78.2. We found a Spearman correlation probabilty of $P<0.001$ between the Balmer decrements and the measured continuum flux ratios, indicating a low probability of no correlation.
We conclude that we are justified in using the continuum color
as a reddening indicator.

In summary, the tests we applied to the data yielded no evidence for more 
obscuration in the S3 sample than in the entire sample.
The interpretation of the test for a color difference
is somewhat muddied by the presence of host
galaxy contamination in some low-luminosity objects.  
For example, one might imagine that the S1 and S2 samples are redder because
they have more galaxy contamination, while the S3 objects are
redder because they are more obscured, leaving the two groups
with the same colors.
Thus colors would not necessarily reveal extra
obscuration in the S3 sample.
To see whether host-galaxy contamination might be affecting
the color test, we looked at the
colors of only the most luminous objects (where $z > 0.45$ and 
$\log L_{5100 \mbox{\AA}} > 43.5$).  
Even in this high luminosity bin, where galaxy
contamination is minimal, the mean color of the S3 sample is similar
to the mean color of the entire sample 
($\log F_{3610}/F_{5100 \mbox{\AA}} = 0.34$ for the S3 sample and
0.35 for the entire sample).
Balmer decrements are difficult to measure if
H$\beta$ is weak.  
Since we have restricted our discussion of the Balmer decrement to those
systems for which the Balmer decrement is reliably measured,
we have excluded exactly those systems with large decrements
that might have cast doubt on color as a measure of reddening.
On the other hand, the number excluded with unreliable measurements of H$\beta$ is only 3\% of the entire sample,
so this is not likely to introduce a large bias.

The previous paragraph shows that one must be cautious not to
overinterpret the tests for excess obscuration in the S3 sample.
We are, nevertheless, left with no convincing evidence for more
obscuration in the S3 sample than in the entire sample,
and must conclude that the greater EW of the [\ion{O}{3}] lines 
in the S3 sample cannot be 
attributed entirely to extra obscuration.  Some additional factor is affecting the equivalent widths. Our comparisons of luminosity and color suggest that obscuration and galaxy contamination is similar in S1 and S3; AGN in S3 may be those in which the strength of [\ion{O}{3}] is sufficient to be further enhanced by obscuration.  Nevertheless we ask what produces this strong [\ion{O}{3}].

\subsection{Spectral Properties}
\label{sec:spec}

Since neither radio-loudness nor orientation can alone account for the
larger NLR EW in the S3 sample, we now turn to a global comparison of
the spectral properties of each subset.  One difference is that the EW$_{\rm{Fe II}}$
appears much larger in the S1 and S2 samples than the S3
sample.
This is visible in the mean spectra for each sample and
also in ALL:PC1, in which \ion{Fe}{2} is anticorrelated with
[\ion{O}{3}].
The \ion{Fe}{2} features are extremely weak or non-existent
in the mean S3 spectrum and all the S3 principal components.
The broad components of the Balmer lines
are still present in the S3 sample, so the BLR is still visible
on average even though the \ion{Fe}{2} lines are not.  For a crude test of the ionization state among the subsets, we used the IRAF routine \textit{splot} to measure the flux ratio of \ion{He}{2} to narrow H$\beta$ from the mean spectra (Figure ~\ref{finalcomp1a}).  We measured the ratio to be 0.432 for S1, 0.309 for S2, and 0.237 for S3.  Thus, it appears that there may be a decrease in ionization state with increasing NLR strength.

To investigate other parameters that may shed light on the physics in the objects in S3, we used K-S tests to compare
the distributions of the entire data set with those of S3.  We compared distributions of
Gauss-Hermite coefficients {\textit{h3}} and \textit{h4}, EW,
and FWHM for H$\beta$ and [\ion{O}{3}] (see Figure ~\ref{hist}).  
With the exception of H$\beta$ \textit{h4} and the [\ion{O}{3}] FWHM the
K-S tests yielded probabilities between 0.35 and 0.99 that the
samples are drawn from the same parent distribution
(0.04 and 0.10 for H$\beta$ \textit{h4} and the [\ion{O}{3}] FWHM respectively).  
We conclude that with the possible exception of  
H$\beta$ \textit{h4} and the [\ion{O}{3}] FWHM the objects
in the S3 sample do not differ significantly from the
entire sample in any of these properties.  A narrower [\ion{O}{3}] FWHM in S3 could imply that the NLR gas is further from the center of the AGN, which would be expected if orientation or covering factor played into the explanation for strong EW$_{\mathrm{[O III]}}$.
An investigation of the
x-ray properties of these objects would be interesting, as x-ray
luminosity has been shown to correlate with $L_{\mathrm{[O III]}}$ \citep{2005ApJ...634..161H, 2006A&A...455..173P} and $\alpha_{ox}$
indicates weaker X-ray flux, compared with the optical, for higher EW$_{\mathrm{[O III]}}$ (BG92), but that is
beyond the scope of this work.

\subsection{Covering Factor}
\label{sec:conc}

Variations in EW$_{\mathrm{[O III]}}$ can also result from
differences in covering factor, density, or a relatively
high ionization parameter.  \cite{2005MNRAS.358.1043B}, hereafter BL05, investigated the
[\ion{O}{3}] line strengths in the PG quasars.  Their
photoionization models suggested that it is covering
factor that causes differences in [\ion{O}{3}].  
If the covering factor is large, 
the NLR gas intercepts a larger amount of continuum and 
produces relatively stronger lines.

To determine whether our strong [\ion{O}{3}] objects show any evidence
of special physical conditions, such as unusual electron density or ionization
parameter, we examined the spectra of the 16 objects in S3 with largest
EW$_{\mathrm{[O III]}}$.  Using Figs 1 and 2 in BL05, and the
[\ion{O}{3}] $\lambda4363/\lambda5007$ and [\ion{O}{3}]
$\lambda5007$/broad H$\beta$ line flux ratios (exhibiting a mean
log~[\ion{O}{3}] $\lambda4363/\lambda5007 = -1.62 \pm 0.10$, and a range of
log~[\ion{O}{3}]~$\lambda5007$/broad H$\beta$ from 1.0 -- 1.18), we
conclude that physical conditions are typical of AGN NLRs
\citep{1997iagn.book.....P}.  
Covering factor then becomes the most likely reason for
the large [\ion{O}{3}] EW in the S3 AGN.
Assuming that we see the true continuum
in the bluest AGN (the presence of \ion{Ca}{2}~K absorption and 
the 4000~\AA\ break in continuum slope show that host
galaxy starlight contributes in the redder AGN, although dust
reddening is also probably involved), we deduce that large but
physical NLR covering factors of $\gtrsim$ 0.35 are required.  However, we cannot make a convincing choice among the various reasons
why the covering factor might be large.
There might be more illuminated gas in the NLR of the S3 objects, perhaps as a result of larger filling factor in the ionization cone, or a larger opening angle.  These two scenarios would result in different infrared properties.  If S3 objects have a larger opening angle, we would expect these objects to have less radiation absorbed and reprocessed into the infrared.  However, if the larger covering factor were due to larger filling factor of the NLR, the infrared properties of S3 objects should be the same as the rest of the data set.

\section{Summary}
\label{sec:summary}
We have presented results of an SPCA of 9046 broad-line AGN from the SDSS DR5 with $0.1 < $z$ < 0.56$ and $10^{42} < L_{5100 \mbox{\AA}} < 10^{46}$ erg s$^{-1}$. We investigated the properties of the rest-frame 4000 - 6000 \AA{}{} spectrum and found that a large range in EW$_{\mathrm{[O III]}}$ dominates the behavior of the sample.  This range causes difficulty in interpreting the PCs.  To try to understand the behavior of [\ion{O}{3}], we divided our sample into three subsets to restrict the broad range of EW$_{\mathrm{[O III]}}$.  Our most extreme [\ion{O}{3}] subset had principal components dominated by NLR emission.  Traditionally, the first principal component in eigenvector analysis (EV1) of quasar spectra has been an anticorrelation between the strengths of \ion{Fe}{2}/H$\beta$ and [\ion{O}{3}]/H$\beta$, and a correlation between the strength of \ion{O}{3}/H$\beta$ with H$\beta$ linewidth.  While the entire sample and subsets show evidence for the anticorrelation between EW$_{\mathrm{[O III]}}$ and \ion{Fe}{2} in their components and mean spectra, we find the H$\beta$ linewidth to be related in only the lowest EW$_{\mathrm{[O III]}}$ subset.  Overall, the width of the Balmer lines seems to have little dependence on the strength of [\ion{O}{3}] or \ion{Fe}{2}.  We find that the differences between our work and BG92 and S03 are caused by the substantially lower luminosities included in our sample.  Also, we find that the usual anticorrelation between EW$_{\mathrm{[O III]}}$ and $L$ only applies for the very high-$L$ end of our sample, and thus the importance of relationships among \ion{Fe}{2}, [\ion{O}{3}], and H$\beta$ linewidth is a strong function of $L$ (or perhaps $L_{\rm{bol}}/L_{\rm{Edd}}$). Our results are consistent with the ``disappearing NLR'' effect suggested by \cite{2004ApJ...614..558N} where high-$L$ AGN experience a decrease in NLR emission due to loss of NLR gas from the system.  

The strong EW$_{\mathrm{[O III]}}$ objects in our sample have dominant NLR lines that do not seem to correlate with any other properties of the broad-line AGN, including $L_{5100 \mbox{\AA}}$, $L_{\rm{bol}}/L_{\rm{Edd}}$, color, BLR properties, or radio emission. We did not find conclusive evidence for the cause of the high EW$_{\mathrm{[O III]}}$ in these objects, although EW$_{\mathrm{[O III]}}$ could be enhanced by suppression of the AGN continuum.  We suggest that the intrinsic range in [\ion{O}{3}] emission is primarily caused by a covering factor that is highly variable among AGN.

We thank Sarah Salviander for access to her data.  Without her cooperation this work would not have been feasible.  Also we acknowledge the helpful comments and discussion provided by Greg Shields and John Barentine.  We thank Todd Boroson for a helpful referee report.


\clearpage

\end{document}